\documentclass[aps,showpacs,nofootinbib]{revtex4}

\usepackage{graphicx}
\usepackage{amsmath}
\usepackage{amssymb}

\newcommand{\be}{\begin{equation}}
\newcommand{\ee}{\end{equation}}
\newcommand{\ba}{\begin{eqnarray}}
\newcommand{\ea}{\end{eqnarray}}

\def\lb{\label}

\begin{document}

\title{ Chiral dynamics in
QED and QCD in a magnetic background and nonlocal noncommutative field 
theories}

\author{E.V. Gorbar}
  \email{egorbar@uwo.ca}
  \altaffiliation[On leave from ]{
       Bogolyubov Institute for Theoretical Physics,
       03143, Kiev, Ukraine}

\author{S. Homayouni}
  \email{shomayou@uwo.ca}

\author{V.A. Miransky}
  \email{vmiransk@uwo.ca}
   \altaffiliation[On leave from ]{
       Bogolyubov Institute for Theoretical Physics,
       03143, Kiev, Ukraine
}
   
\affiliation{
Department of Applied Mathematics, University of Western
Ontario, London, Ontario N6A 5B7, Canada
}

\date{\today}

\begin{abstract}
We study the connection of the chiral dynamics in QED and QCD
in a strong magnetic field
with noncommutative field theories (NCFT).
It is shown
that these 
dynamics determine complicated nonlocal NCFT. In particular, 
although the 
interaction vertices for electrically neutral composites
in these gauge models can be represented
in the space with noncommutative spatial coordinates, there is
no field transformation that could put the vertices in the 
conventional form considered in the literature. 
It is unlike the Nambu-Jona-Lasinio (NJL) model
in a magnetic field where such a field transformation can be found,
with a cost of introducing an exponentially damping form factor
in field propagators.
The crucial distinction between these two types of models is
in the characters of their interactions, being short-range in 
the NJL-like models
and long-range in gauge theories.
The relevance of the NCFT connected with the gauge
models
for the description of the
quantum Hall effect in condensed matter systems with long-range
interactions is briefly discussed.
\end{abstract}


\maketitle

\section{Introduction}
\label{1}

During last few years, different aspects of
noncommutative field theories (NCFT) have been intensively studied
(for reviews, see Ref. \cite{DNS}). In particular,
it was revealed that NCFT are intimately related to the dynamics in
quantum mechanical models in a strong magnetic field \cite{DJT,BS},
nonrelativistic field systems in a strong magnetic field 
\cite{IKS,CTZ},
nonrelativistic magnetohydrodynamical field theory \cite{GJPP},
and, in the case of open strings
attached to $D$-branes, to the dynamics in string theories in
magnetic backgrounds \cite{strings,SW}. 

Recently, the connection between the dynamics
in relativistic field theories
in a strong homogeneous magnetic field and that in NCFT has been studied 
\cite{GM}.
The main conclusion of that paper
was that although relativistic field theories in the regime with the
lowest Landau level (LLL) dominance
indeed determine a class
of NCFT, these NCFT are different from
the conventional ones considered in the literature. In particular,
the UV/IR mixing, taking place in the
conventional NCFT \cite{MRS}, is absent in these theories. The reason of 
that is an inner structure (i.e., dynamical form factors) of 
electrically neutral
composites in these theories. We emphasize that in order to establish
the connection between dynamics in a homogeneous magnetic field and 
dynamics in
NCFT, it
is necessary to consider neutral fields. The point is that 
in 
homogeneous magnetic backgrounds, momentum is a good quantum number only 
for neutral
states and therefore 
one can introduce
asymptotic states and S-matrix only for them.

While studies of the origins of the
noncommutativity in relativistic 
quantum field theories in a magnetic field are interesting 
in themselves, it is even more important that they lead to new physical
results. In particular, as was shown in Ref. \cite{GM}, the NCFT approach 
allows to derive interaction vertices for neutral composites. 
These vertices
automatically include (in the form of the Moyal product \cite{DNS})
{\it all} powers of transverse derivatives [i.e., the derivatives 
with respect to coordinates orthogonal to a magnetic field]. 
This result is quite noticeable because the dimension of the
transverse subspace is two and it is 
very seldom 
that one can with a good accuracy
calculate vertices for composites 
in quantum field
models with spatial dimensions higher than one. 

In the analysis in Ref. \cite{GM}, the Nambu-Jona-Lasinio (NJL) model
in a magnetic field was considered. 
It was shown that there exist two equivalent 
descriptions of its dynamics. In the first description, one uses the 
conventional
composite operators $\sigma(x) \sim \bar{\psi}(x) \psi (x)$
and $\pi(x) \sim i\bar{\psi}(x)\gamma_5 \psi (x)$. In this case,
besides the usual
Moyal factor, the
additional Gaussian-like (form-)factor
$e^{-\left(\sum_{i=1}^n\vec{k}_{\perp i}^2\right)/{4|eB|}}$
occurs in n-point interaction vertices of the fields $\sigma(x)$ 
and $\pi(x)$. Here $\vec{k}_{\perp i}$ is a momentum 
of the $i$-th composite in a plane
orthogonal to the magnetic field. 
These form factors reflect an inner
structure of composites and play an important role in providing
consistency of these NCFT. In particular, because of them,
the UV/IR mixing is absent in these theories. In the second
description, one
considers other, ``smeared", fields $\Sigma(x)$ and $\Pi(x)$,
connected with $\sigma(x)$ and  $\pi(x)$ through a non-local
transformation. Then, while the additional factors are removed in the
vertices for the smeared fields, they appear in their propagators,
again resulting in the UV/IR mixing removal. 

The Gaussian form of 
the exponentially damping form factors reflects the
Landau wave functions of fermions on the LLL. 
The form factors are intimately 
connected with the holomorphic representation in the problem of
quantum
oscillator (for a review of the holomorphic 
representation, see Ref. \cite{FS}). Indeed,   
in the problem of a free fermion in a magnetic field,
the dynamics in a plane orthogonal to the magnetic field
is an
oscillator-like one.
\footnote {In particular, the holomorphic 
representation 
is widely
used for the description of the quantum Hall effect \cite{GK}.} 
And because weak short-range interactions between fermions in the
NJL model in a magnetic field
do not change this feature of
the dynamics, the form factors in that model
have the Gaussian form. But what happens
in the case of more sophisticated dynamics, such as those with
long-range interactions  in gauge models? To find the answer 
to this question is the primary goal of the present paper.  

In this work, we will extend the analysis of Ref. \cite{GM}
to the more complicated
cases of QED and QCD in a strong magnetic field. It will be shown 
that in these gauge models, the connection of the dynamics with NCFT 
is much more 
sophisticated.
It is not just that the damping form-factors are not
Gaussian in these models
but there does not exist an analogue of the smeared fields at all. 
As a result, their interaction vertices cannot be 
transformed into the form of vertices in conventional NCFT. On the other
hand, it is quite remarkable that, by using the Weyl symbols of the
fields \cite{DNS}, 
their vertices 
can nevertheless be represented  
in the space with noncommutative spatial coordinates. The dynamics
they describe correspond to complicated nonlocal NCFT. We will call these 
theories type II nonlocal NCFT. The name type I nonlocal NCFT will be
reserved for models similar to the NJL model in a magnetic field, for 
which
smeared fields exist.
In both these cases, the term ``nonlocal"  
reflects the point that, 
besides the Moyal factor, additional form factors 
are present in the theories.

The crucial distinction between these two types of models is
in the characters of their interactions. While the interaction in the 
NJL-like
models is local (short-range), it is long-range in gauge theories. 
This point is reflected in a much richer structure of neutral composites
in the latter. We believe that both these types of
nonlocal NCFT can be relevant not only for relativistic field theories
but also for nonrelativistic systems in a magnetic field.
In particular, while type I NCFT can be relevant for the
description of the quantum Hall effect in condensed matter systems
with short-range interactions \cite{IKS,CTZ,Hall}, 
type II NCFT can be relevant in studies of this effect in
condensed 
matter systems with long-range interactions (such as carbon materials).   

The paper is organized as follows. 
In Sec. \ref{catalysis}, a brief review of the dynamics of spontaneous
chiral symmetry breaking 
in QED and QCD in a magnetic field 
is made.
In Sec. \ref{2}, by using the
formalism of bilocal composite fields,
the effective action and interaction vertices for Nambu-Goldstone
composites are derived in QED in a strong magnetic field. In Sec.
\ref{types}, these vertices are written in the space with noncommutative
spatial coordinates and the concept of nonlocal type I and type II
NCFT is introduced. In Sec. \ref{NJL}, the dynamics in QED with
a large number of fermion flavors $N_f$ in a magnetic field is
studied in a special dynamical regime with local (short-range) 
interactions.
In this regime, the results obtained in the NJL model in a magnetic
field \cite{GM} are reproduced in the formalism of bilocal composite
fields. In Sec. \ref{QED}, it is shown that
the chiral dynamics in QED with a weak coupling
in a strong magnetic field is related to type II nonlocal NCFT.
In Sec. \ref{QCD}, it is shown that the chiral dynamics in
QCD in a strong magnetic field also relates to such a type of nonlocal
NCFT. In Section \ref{general}, we discuss the origins of the
connection between relativistic field theories in magnetic 
backgrounds and NCFT. 
In Section \ref{conclusion}, we summarize the main
results of the paper and discuss their possible applications.  
In the Appendix, some useful
formulas and relations are derived.

\section{Chiral symmetry breaking in QED and QCD in a 
magnetic field}
\label{catalysis} 

The central dynamical
phenomenon in relativistic field theories in a magnetic
field 
is the phenomenon of the magnetic catalysis:
a constant magnetic field is a strong catalyst of dynamical chiral
symmetry breaking, leading to the generation of a fermion dynamical mass
even at the weakest attractive interaction between fermions
\cite{GMS1,GMS2}. The essence of this effect is the dimensional
reduction in the dynamics of fermion pairing in a strong magnetic
field, when the LLL dynamics dominates. In this section, we will
indicate those features 
of this phenomenon in QED and QCD which are relevant for the present work. 

The consistent theory of this phenomenon in QED was developed in
Refs. \cite{QED1}  
[for earlier papers considering spontaneous chiral symmetry breaking in
QED in a magnetic field, see Refs. 
\cite{GMS2,QED2}]. The crucial point of the analysis \cite{QED1}
was to recognize that there exists a special non-local (and non-covariant)
gauge in which the so called improved rainbow (ladder) approximation is 
reliable in 
this problem, i.e., there is a consistent truncation 
of the system of the Schwinger--Dyson equations.
\footnote {We recall that in the improved rainbow (ladder)
approximation the vertex $\Gamma^{\nu}(x,y,z)$ is taken to be bare and the
photon propagator is taken in the one-loop approximation.}
The full
photon propagator in this gauge has the form
\begin{equation}
{D}_{\mu\nu}(k)=
i\frac{g^{\parallel}_{\mu\nu}}{k^2+k_{\parallel}^2\,
\Pi(k_{\perp}^2, k_{\parallel}^2)} 
+ \,i\frac{g^{\perp}_{\mu\nu}k^2 
- (k^{\perp}_{\mu}k^{\perp}_{\nu} + k^{\perp}_{\mu}
k^{\parallel}_{\nu} + k^{\parallel}_{\mu}k^{\perp}_{\nu})}
{(k^2)^2},
\label{non-l}
\end{equation}
where $\Pi(k_{\perp}^2, k_{\parallel}^2)$ is the polarization 
operator and the symbols $\perp$ and $\parallel$ in $g_{\mu\nu}$ 
and $k_\mu$ are
related to the transverse
$(1,2)$ and longitudinal $(0,3)$ space-time components, respectively
(we consider a constant magnetic field $B$ directed in the $+x_3$
direction). Since the transverse 
degrees of freedom decouple from the LLL dynamics \cite{QED1}, 
\footnote {This feature reflects the point that the spin of the
LLL fermion (antifermion) states is polarized along (opposite to)
the magnetic field.}
only the
first term in photon propagator $D_{\mu\nu}$ (\ref{non-l}), 
proportional to
$g^{\parallel}_{\mu\nu}$, is relevant. 
Therefore,
as the full photon propagator in this
special gauge, one can take the Feynman-like
noncovariant propagator
\begin{equation}
{D}_{\mu\nu}(k)=
i\frac{g^{\parallel}_{\mu\nu}}{k^2+k_{\parallel}^2
\Pi(k_{\perp}^2, k_{\parallel}^2)}.
\label{trunc}
\end{equation}
It is important that this propagator does 
not
lead to infrared mass singularities in loop corrections in a
vertex \cite{QED1} that makes the improved rainbow approximation
to be reliable in this gauge (because of 
mass singularities in covariant gauges, the loop corrections in the vertex 
are large there \cite{GMS2,QED1}).

It was shown in Ref. \cite{QED1} that the kinematic region mostly 
responsible
for generating the fermion mass is that
with $m_{dyn}^2 \ll |k_{\parallel}^2|\,,\, k_{\perp}^2 \ll |eB|$.
In that region, fermions can be
treated as massless and, as a result,
the polarization operator can be calculated in one-loop approximation.
It is
\ba
\Pi(k_{\perp}^2,k_{\parallel}^2)
\simeq -\frac{2\tilde{\alpha_b} |eB|}{\pi k_{\parallel}^2},
\lb{polarization}
\ea
where $\tilde{\alpha}_b \equiv N_f\alpha_b=\frac{N_{f}e_{b}^2}{4\pi}$.
Here $N_f$ is the number of fermion flavors
and $\alpha_b$ is the QED running coupling
related to the magnetic scale $\mu = \sqrt{|eB|}$.

Thus, in this
approximation, photon propagator (\ref{trunc})
becomes a propagator of a free
massive boson with $M_{\gamma}^2=2\tilde{\alpha_b}|eB|/\pi$:
\begin{equation}
{D}_{\mu\nu}(x)=
\frac{i}{(2\pi)^4} \int d^4k\, e^{-ikx}\,
\frac{g^{\parallel}_{\mu\nu}}{k^2 - M_{\gamma}^2}.
\label{trunc1}
\end{equation}

As was shown in 
the first two papers of Ref. \cite{QED1}, the improved rainbow 
approximation
with the photon propagator (\ref{trunc}) is reliable when the 
parameter $\tilde{\alpha}_b$ is 
small, 
i.e.,
$\tilde{\alpha}_b \ll 1$. 
In the case of large $N_f$, when one can use the $1/N_f$ expansion,
this approximation is reliable for 
arbitrary $\tilde{\alpha}_b$  
[see the last paper in Ref. \cite{QED1}].

In the weak coupling regime with
$\tilde{\alpha_b} =  N_f\alpha_b \ll 1$,
the dynamically generated mass of fermions is:
\ba
m_{dyn} = C |eB|^{1/2} F(\tilde{\alpha_b}) \exp\left[-\frac{\pi N_f}
{\tilde{\alpha_b}\ln(C_1/\tilde{\alpha_b})}\right],
\lb{mass}
\ea
where $F(\tilde{\alpha_b}) \simeq  (\tilde{\alpha_b})^{1/3}$,
$C_1 \simeq 1.82$ and
$C$ is a numerical constant of order one. In the strong coupling
limit for large $N_f$, the dynamical mass takes the form 
\ba
m_{dyn} \simeq \sqrt{|eB|} \mbox{exp}(-N_f).
\lb{mass1}
\ea

With appropriate modifications, this theory was extended to the
case of QCD in a strong magnetic field in Ref. \cite{QCD1}
[for earlier consideration of QCD in a magnetic field, see Refs. 
\cite{QCD2,KLW}]. These modifications will be described in Sec. \ref{QCD}.
The results of 
Refs. \cite{GMS2,QED1} and \cite{QCD1} will be an essential
ingredient in the analysis in this paper.

\section{QED in a magnetic field: The effective action }
\label{2}

In this Section, we analyze the dynamics in the
chiral symmetric QED in a strong magnetic field. 
We will consider both  
the weakly
coupling regime, with $\tilde{\alpha}_b \ll 1$, and (for large $N_f$)
the strongly
coupling regime with
$\tilde{\alpha}_b \gtrsim 1$. In both these cases, one can use the
results of the analysis of Ref. \cite{QED1}.

The chiral symmetry in this model is
$SU(N_f)_{L} \times SU(N_f)_{R}$. The generation of a fermion mass
breaks this symmetry down to $SU(N_f)_{V}$ and, as
a result, $N_{f}^2 - 1$ neutral Nambu-Goldstone (NG) composites
$\pi^A$, $A = 1, 2,...,N_{f}^2 - 1$, occur (we do not consider here
the anomalous $U(1)_A$). 
Our aim in this section is to
derive the interaction vertices for $\pi^A$ in the regime with
the LLL dominance and clarify whether
their structure corresponds to a NCFT.

Integrating out the photon field $A_{\mu}$, we obtain the following 
non-local effective action for fermions in QED in a magnetic field:
\ba
S = \int d^4x\, \bar{\psi}i\gamma^{\mu}{\cal D}_{\mu}\psi  -
2i\pi \alpha_{b}\int d^4xd^4y\, 
\bar{\psi}(x)\gamma^{\mu}\psi(x)
D_{\mu\nu}^{(0)}(x-y)\bar{\psi}(y)\gamma^{\nu}\psi(y),
\label{initialaction}
\ea
where, in the lowest order, the bare propagator corresponding to
propagator (\ref{trunc}) is
$$
D_{\mu\nu}^{(0)}(x-y) = \frac{i}{(2\pi)^4} \int d^4k\, e^{-ik(x - y)}\,
\frac{g^{\parallel}_{\mu\nu}}{k^2},
\label{bare}
$$
and the vector potential $A_{\mu}^{ext}$ in the covariant derivative
${\cal D}_{\mu} = \partial_{\mu} - ieA_{\mu}^{ext}$
in (\ref{initialaction}) describes a constant
magnetic field $B$ directed in the $+x^3$ direction. In the symmetric
gauge, it is
$A_{\mu}^{ext} = (0,\frac{Bx^2}{2},\frac{-Bx^1}{2},0)$.
Following the auxiliary field method developed for theories with
nonlocal interaction in Refs. \cite{Kl,K}, we add the term
\ba
\Delta S = - 2i\pi \alpha_{b}\int d^4xd^4y\, 
\mbox{tr}\left\{\gamma^{\mu}\,
[\varphi_{a}^{b}(x,y)-
\psi_{a}(x)\bar{\psi}^{b}(y)]\,\gamma^{\nu}\,[\varphi_{b}^{a}(y,x) - 
\psi_{b}(y)\bar{\psi}^a(x)]
\right\}D_{\mu\nu}^{(0)}(x-y)
\lb{auxil}
\ea
in the action. Here $\varphi_{a}^{b}(x,y)$ is a bilocal auxiliary field 
with
the indices $a$ and $b$ from the fundamental representation of
$SU(N_f)$. Then we 
obtain the action             
$$
S = \int d^4x\, \bar{\psi}i\gamma^{\mu}{\cal D}_{\mu}\psi  -
4i\pi \alpha_{b}
\int d^4xd^4y\, \bar{\psi}(x)\gamma^{\mu}\varphi(x,y)\gamma^{\nu}\psi(y)
D_{\mu\nu}^{(0)}(x-y) 
$$
\ba
-2i\pi \alpha_{b}\int d^4xd^4y\, \mbox{tr}\,[\gamma^{\mu}\,
\varphi(x,y)\,\gamma^{\nu}\,\varphi(y,x)]\, D_{\mu\nu}^{(0)}(x-y)
\label{intermediateaction}
\ea
[here, for clarity of the presentation, we omitted the $SU(N_f)$
indices]. 
Integrating over fermions, we find
$$
S(\varphi) = -i\mbox{TrLn}\left[\gamma^{\mu}i{\cal D}_{\mu}\delta^4(x-y) -
4i\pi \alpha_{b} \gamma^{\mu}\varphi(x,y)\gamma^{\nu} 
D_{\mu\nu}^{(0)}(x-y)\right] 
$$
\ba
-2i\pi \alpha_{b} \int d^4xd^4y\, \mbox{tr}[\gamma^{\mu}  
\varphi(x,y)\gamma^{\nu}\varphi(y,x)]
D_{\mu\nu}^{(0)}(x-y),
\lb{action}
\ea
where 
$\mbox{Tr}$ and $\mbox{Ln}$ are taken in the functional sense.

Following Ref. \cite{K}, we can expand $\varphi(x,y)$ as
\ba
\varphi(x,y)= \varphi_0(x,y) + \tilde{\varphi}(x,y),
\lb{exp}
\ea
\ba
\tilde{\varphi}(x,y) =
\sum_n \int \frac{d^4P}{(2\pi)^4} \phi_n(P)\chi^{(l)}_n(x,y;P).
\lb{expansion}
\ea
Here $\varphi_0(x,y)$ satisfies the equation
\ba
\frac{\delta S}{\delta \varphi} = 0,
\lb{extremum}
\ea
which is equivalent to the Schwinger--Dyson
equation
\ba
S_{(l)}^{-1}(x,y) = S^{-1}_0(x,y) -
4\pi \alpha_b \gamma^{\mu}S_{(l)}(x,y)\gamma^{\nu} D_{\mu\nu}^{(0)}(x-y),
\lb{SDequation}
\ea
where $S_0$ is the bare fermion propagator and $S_{(l)} \equiv \varphi_0$
is the full fermion propagator in the rainbow (ladder) approximation.
As to Eq.(\ref{expansion}), $\phi_n(P)$ is a field 
operator describing a neutral composite $|n,P>$
and  
$\chi^{(l)}_n(x,y;P)$ are solutions of the {\it off-mass-shell} 
Bethe--Salpeter 
(BS) equation in the ladder approximation,
\ba
\chi^{(l)}(x,y;P) = 4\pi \alpha_{b} \lambda(P) \int d^4x_1d^4y_1\, 
S_{(l)}(x,x_1)\,\gamma^{\mu}
\chi^{(l)}(x_1,y_1;P)\,\gamma^{\nu}S_{(l)}(y_1,y)\,
D_{\mu\nu}^{(0)}(x_1-y_1).
\lb{BSequation}
\ea
The insertion of
factor $\lambda(P) \neq  1$ in this equation
allows to consider off-mass-shell states
with an arbitrary mass $M^2 = P^2$. 
The on-mass-shell states correspond to $\lambda(P) = 1$.

Using Eqs. (\ref{exp}) and (\ref{SDequation}), the action
(\ref{action}) can be rewritten as 
$$
S(\tilde{\varphi}) = -i\mbox{TrLn}\left[S_{(l)}^{-1}(x,y) -
4\pi \alpha_{b}
\gamma^{\mu}\tilde{\varphi}(x,y)\gamma^{\nu} D_{\mu\nu}^{(0)}(x-y)\right]
$$
\ba
-2i\pi \alpha_{b}\int d^4xd^4y\, \mbox{tr}[\gamma^{\mu}
(\varphi_0(x,y) + \tilde{\varphi}(x,y))\gamma^{\nu}
(\varphi_0(y,x) + \tilde{\varphi}(y,x))] D_{\mu\nu}^{(0)}(x-y).
\lb{action1}
\ea
Expanding now the action $S(\tilde{\varphi})$ in powers of
$\tilde{\varphi}$ and ignoring its part that does not depend on
$\tilde{\varphi}$, we obtain
$$
S(\tilde{\varphi}) = \sum_{n=2}^{\infty} \frac{i}{n} \int 
d^4x_1d^4y_1\,...\,d^4x_n
d^4y_n\,
\mbox{tr}\,[S_{(l)}(x_1,y_1)\varphi_D(y_1,x_2)S_{(l)}(x_2,y_2)
\varphi_D(y_2,x_3)
\,.\,.\,.\,S_{(l)}(x_{n-1},y_n)\varphi_D(y_n,x_1)\,]
$$
\ba
- 2i\pi \alpha_{b}\int d^4xd^4y \mbox{tr}[\gamma^{\mu}
\tilde{\varphi}(x,y)\gamma^{\nu}\tilde{\varphi}(y,x)] D_{\mu\nu}^{(0)}(x-y),
\lb{action011}
\ea
where
$$
\varphi_D(x,y) = 4\pi \alpha_{b} 
\gamma^{\mu}\tilde{\varphi}(x,y)\gamma^{\nu}D_{\mu\nu}^{(0)}(x-y).
$$
Because $\varphi_0$ satisfies the Schwinger--Dyson equation (\ref{extremum}),
the term linear in $\tilde{\varphi}$ is absent
in (\ref{action011}).

As is clear from the discussion in the previous section, 
one should use the improved rainbow (ladder) approximation
in the present problem.
The Schwinger--Dyson equation for the fermion
propagator in this approximation takes the form
\ba
S^{-1}(x,y) = S^{-1}_0(x,y) - 4\pi \alpha_b \gamma^{\mu}S(x,y)\gamma^{\nu}
D_{\mu\nu}(x-y),
\lb{SDequation1}
\ea
where the photon propagator $D_{\mu\nu}(x)$ is given in Eq. 
(\ref{trunc1}).
The off-mass-shell BS equation
in the improved ladder approximation is given by
\ba
\chi(x,y;P) = 4\pi \alpha_{b} \lambda(P) \int d^4x_1d^4y_1\, 
S(x,x_1)\,\gamma^{\mu}
\chi(x_1,y_1;P)\,\gamma^{\nu}S(y_1,y)\,D_{\mu\nu}(x_1-y_1).
\lb{BSequation1}
\ea
The comparison of Eqs. (\ref{SDequation1}), (\ref{BSequation1}) 
with Eqs. (\ref{SDequation}),
(\ref{BSequation}) suggests that in the improved 
rainbow (ladder) approximation
the effective action (\ref{action011}) should be
replaced by the following one:
$$
S(\tilde{\varphi}) = \sum_{n=2}^{\infty} \frac{i}{n} \int 
d^4x_1d^4y_1\,...\,d^4x_n
d^4y_n\,
\mbox{tr}\,[S(x_1,y_1)\varphi_D(y_1,x_2)S(x_2,y_2)
\varphi_D(y_2,x_3)
\,.\,.\,.\,S(x_{n-1},y_n)\varphi_D(y_n,x_1)\,]
$$
\ba
- 2i\pi \alpha_{b}\int d^4xd^4y \mbox{tr}[\gamma^{\mu}
\tilde{\varphi}(x,y)\gamma^{\nu}\tilde{\varphi}(y,x)] D_{\mu\nu}(x-y),
\lb{action11}
\ea
where
$$
\varphi_D(x,y) = 4\pi \alpha_{b} 
\gamma^{\mu}\tilde{\varphi}(x,y)\gamma^{\nu}D_{\mu\nu}(x-y)
$$
and 
\ba
\tilde{\varphi}(x,y)=\sum_n \int \frac{d^4P}{(2\pi)^4} 
\phi_n(P)\chi_n(x,y;P)
\lb{expansion1}
\ea
(compare with Eq. (\ref{expansion})). Here $\chi_n(x,y;P)$ are 
solutions of the off-mass-shell BS equation (\ref{BSequation1}).

As in the case of the NJL model \cite{GM}, in the improved rainbow
approximation in QED,
the LLL fermion propagator
factorizes into two parts: the part depending on the
transverse coordinates $x_{\perp} = (x^1,x^2)$ and that depending on the
longitudinal coordinates $x_{||} = (x^0,x^3)$,
\ba
S_{LLL}(x,y) = P(x_{\perp},y_{\perp})S_{||}(x_{||}-y_{||}).
\lb{LLLpropagator}
\ea
Here $P(x_{\perp},y_{\perp})$ is the projection operator on the LLL states
\cite{GM}
which in the symmetric gauge is
\ba
P(x_{\perp},y_{\perp}) = \frac{|eB|}{2\pi}
e^{\frac{ieB}{2}\epsilon^{ab}x^ay^b}
e^{-\frac{|eB|}{4}(\vec{x}_{\perp}-\vec{y}_{\perp})^2}.
\lb{projector}
\ea
The first exponential factor in $P(x_{\perp},y_{\perp})$ is the
Schwinger phase \cite{Schwinger}. Its presence is dictated by the
group of magnetic translations in this problem (for more details,
see Sec. \ref{general} below).
As was shown in Ref. \cite{GM},
it is the Schwinger phase that is responsible for producing the Moyal
factor
(a signature of NCFT \cite{DNS})
in interaction vertices.

As to the longitudinal part, in the improved rainbow approximation
it has the form \cite{QED1}
\ba
S_{\|}(x_{\|}-y_{\|}) = \int \frac{d^2k_{\|}}{(2\pi)^2}
e^{ik_{\|}(x^{\|}-y^{\|})} \frac{i}{k_{\|}\gamma^{\|} - m(k_{\|}^2)}\,
\frac{1 - i\gamma^1\gamma^2}{2},
\lb{flatspace}
\ea
i.e., it has the form of a fermion propagator in 1+1 dimensions.
\footnote {In particular,
the matrix $(1 - i\gamma^1\gamma^2)/2$ is the
projection operator on the fermion (antifermion) states with the spin
polarized along (opposite to) the magnetic field, and therefore it
projects on two states of the four ones, as should be in 1+1
dimensions.}
The dynamical mass function $m(k_{\|}^2)$ is
essentially constant for
$k_{\parallel}^2 \alt  |eB|$ and
rapidly decreases for $k_{\parallel}^2 > |eB|$ \cite{QED1}.
Therefore, a simple and reliable approximation for $m(k_{\|}^2)$
is
\ba
m(k_{\|}^2) = \theta (|eB| - k_{\parallel}^2)\, m_{dyn},
\lb{mfunction}
\ea
where $\theta (x)$ is the step function and $m_{dyn}$ is the fermion pole
mass
(\ref{mass}) or (\ref{mass1}) [this conclusion was later confirmed
in papers \cite{QED3}].

The operators  $\phi_n(P)$ in equation (\ref{expansion1}) describe all
possible neutral fermion-antifermion composites.
The description of their interaction vertices
in QED in a magnetic field is quite a formidable problem.
Henceforth we limit ourselves to considering only the interaction
vertices for the NG boson states $|A;P>$ and their operators
$\phi^{A}(P)$ (for a discussion concerning other states, see
Sec. \ref{general}).
\footnote {As is well known, in some approximations, the BS 
equation
is plagued by the appearance of spurious solutions. Because we restrict
ourselves to calculating the vertices for the NG bosons, which are
manifestly physical, no such problem occurs in this study.}

In a magnetic field,
the wave function of the states $|A;P>$ satisfying BS
equation (\ref{BSequation1})
has the following form \cite{GMS2}:
\ba
\chi^{A}(x,y;P) \equiv <0|T\psi(x)\bar{\psi}(y)|A;P> = e^{-iPX}
e^{ier^{\mu}A_{\mu}^{ext}(X)} \tilde{\chi}^{A}(r;P),
\lb{chitilde}
\ea
where $r=x-y$, $X=\frac{x+y}{2}$ and, that is very important,
the function $\tilde{\chi}^{A}(r;P)$ is independent of 
the center of mass coordinate $X$.
This fact reflects the existence of the group of 
magnetic translations in the present problem. 
As in the case of the fermion propagator, the presence of
the Schwinger factor $e^{ier^{\mu}A_{\mu}^{ext}(X)}$
in expression (\ref{chitilde}) is dictated by this symmetry
(see Sec. \ref{general} below).
As to the $SU(N_f)$ structure of $\tilde{\chi}^{A}(r;P)$, it is:
\ba
\tilde{\chi}^{A}(r;P) = \frac{\lambda^{A}}{2} \tilde{\chi}(r;P),
\lb{chitilde1}
\ea
where $\lambda^{A}$ are $N_{f}^2 - 1$ matrices in the fundamental
representation of $SU(N_f)$.

Now, transforming BS equation (\ref{BSequation1})
into momentum space, we get:
$$
\tilde{\chi}^{A}(p;P) = \frac{16\pi \alpha_{b}
\lambda(P)}{|eB|^2} \int
\frac{d^2q_{\perp}d^2A_{\perp}d^2k_{\perp}d^2k_{||}}{(2\pi)^6}
e^{i(P_{\perp}-q_{\perp}) \times (A_{\perp}-p_{\perp})}
e^{-\frac{(\vec{p}_{\perp}+\vec{A}_{\perp})^2}{2|eB|}}
e^{-\frac{\vec{q}_{\perp}^2}{2|eB|}}
$$
\ba
\times\,\, S_{||}(p_{||}+\frac{P_{||}}{2})
\gamma^{\mu} \tilde{\chi}^{A}(k;P)\gamma^{\nu} 
S_{||}(p_{||}-\frac{P_{||}}{2})
D_{\mu\nu}(k_{||}-p_{||},\vec{k}_{\perp}-\vec{A}_{\perp}),
\lb{BSmomentum}
\ea
where $p_{\perp} \times q_{\perp} \equiv \frac{\epsilon^{ab}p^aq^b}{eB}$
is the Moyal cross product.
Then, introducing the variable $\vec{u}_{\perp}= \vec{A}_{\perp}-
\vec{k}_{\perp}$ and representing $\tilde{\chi}^{A}$ as
\ba
\tilde{\chi}^{A}(p;P) = e^{-\frac{\vec{p}_{\perp}^2}{|eB|}} e^{-iP_{\perp} 
\times p_{\perp}} f^{A}(p;P),
\lb{tildechi}
\ea
we integrate over $q_{\perp}$ in (\ref{BSmomentum}) and find the following 
equation for the function $f^{A}(p;P)$:
$$
f^{A}(p;P) = \frac{8 \alpha_{b}\lambda(P)}{|eB|} \int 
\frac{d^2u_{\perp}d^2k_{\perp}d^2k_{||}}{(2\pi)^4}\,
e^{iP_{\perp} \times u_{\perp}}
e^{-\frac{(\vec{u}_{\perp} + \vec{k}_{\perp})^2}{|eB|}}
e^{-\frac{\vec{k}_{\perp}^2}{|eB|}}\,
$$
\ba
\times\,\, S_{||}(p_{||}+\frac{P_{||}}{2})
\gamma^{\mu} f^{A}(k;P)\gamma^{\nu} S_{||}(p_{||}-\frac{P_{||}}{2})
D_{\mu\nu}(k_{||}-p_{||},\vec{u}_{\perp}).
\lb{f-equation1}
\ea

Because the right hand side of equation (\ref{f-equation1}) does not 
contain
$p_{\perp}$, we conclude that $f^{A}(p;P)$ does not depend on $p_{\perp}$, 
i.e., it
is a function of $p_{||}$ and $P$ only.  
[It is not difficult to
convince yourself that this fact is a
direct consequence of the factorization of the LLL propagator.]
Therefore we
can explicitly integrate over $k_{\perp}$ and get
\ba
f^{A}(p_{||};P) = 4\pi \alpha_{b}\lambda(P) \int 
\frac{d^2u_{\perp}d^2k_{||}}{(2\pi)^4}
e^{iP_{\perp} \times u_{\perp}}
e^{-\frac{\vec{u}_{\perp}^2}{2|eB|}}
S_{||}(p_{||}+\frac{P_{||}}{2})
\gamma^{\mu} f^{A}(k_{||};P)\gamma^{\nu} S_{||}(p_{||}-\frac{P_{||}}{2})
D_{\mu\nu}(k_{||}-p_{||},\vec{u}_{\perp}).
\lb{f-equation2}
\ea

Henceforth, in the main body of the paper, we will consider
the dynamics for the case of zero longitudinal
momenta, $P_{||}=0$, i.e., for
$\phi^{A}(P) = (2\pi)^2\delta^2(P_{||})\phi^{A}(P_{\perp})$, 
when the function $\tilde{\varphi}(x,y)$ in Eq. (\ref{expansion1})
is
\ba
\tilde{\varphi}(x,y)= \int \frac{d^2P_{\perp}}{(2\pi)^2} \phi^{A}(P_{\perp})
\chi^{A}(x,y;P_{\perp})
\lb{expansion2}
\ea
with the BS wave function $\chi^{A}$
depending only on $P_{\perp}$. 
For $P_{||}=0$,
the Fourier transform of the fields $\phi^{A}$ 
in coordinate space depends only on
the coordinates $X_{\perp}$. 
Because in this problem the noncommutative geometry is 
connected
only with the transverse coordinates, this dependence is
the most relevant for our purposes. The case with nonzero
longitudinal momenta $P_{||}$ is quite cumbersome and is
considered in the Appendix.

We would like to 
emphasize that in
the case of nonlocal interactions, such as those in QED, 
the bound state problem with nonzero $P$
is quite formidable. 
In 
fact, as will become clear below (see also Appendix),
it is the very special property of
the factorization of the longitudinal and transverse dynamics 
in the LLL 
approximation that 
will allow us to succeed in the 
derivation of explicit expressions for interaction
vertices for NG fields $\phi^{A}$ 
for nonzero $P$.  

When $P_{||} = 0$, one can check that, up to the
factors $\lambda(P_{\perp})$ and $e^{iP_{\perp} \times u_{\perp}}$,
the structure of equation (\ref{f-equation2}) is similar
to the BS equation for NG bosons with 
$P_{\perp} = P_{||} = 0$ considered 
in Ref. \cite{GMS2}. By
using the analysis performed in that work, we immediately find that
\ba
f^{A}(p_{||};P_{\perp}) = S_{||}(p_{||})F^{A}(p_{||};P_{\perp})\gamma^5
\frac{1-i\gamma^1\gamma^2}{2}S_{||}(p_{||}),
\lb{pion}
\ea
where $F^{A}(p_{||};P_{\perp})$ is a scalar function. It  
satisfies the 
following
equation in Euclidean space:
\ba
F^{A}(p_{||};P_{\perp}) = 8\pi \alpha_{b} \lambda(P_{\perp}) \int 
\frac{d^2u_{\perp}d^2k_{||}}{(2\pi)^4}
\frac{F^{A}(k_{||};P_{\perp})}{k_{||}^2+m^2}
\frac{e^{iP_{\perp} \times u_{\perp}} e^{-\frac{\vec{u}_{\perp}^2}{2|eB|}}}
{(k_{||}-p_{||})^2+\vec{u}_{\perp}^2 + M^{2}_{\gamma}}.
\lb{Aequation}
\ea
In the derivation of this equation, Eq. (\ref{trunc1}) was used.

Now, taking into account Eqs.(\ref{chitilde}), (\ref{tildechi}),
and (\ref{pion}), and integrating explicitly over $p_{\perp}$, 
we find that the BS wave function in coordinate space is
$$
\chi^A(x,y;P_{\perp})= P(x_{\perp},y_{\perp}) \int \frac{d^2p_{||}}
{2(2\pi)^2} e^{i\vec{P}_{\perp}\frac{\vec{x}_{\perp}+\vec{y}_{\perp}}{2}}
e^{-ip_{||}(x^{||}- y^{||})}
e^{-\frac{\vec{P}_{\perp}^2}{4|eB|}}
e^{\frac{\epsilon^{ab}P_{\perp}^a(x_{\perp}^b-y_{\perp}^b) \mbox{sign}(eB)}{2}}
\,
$$
\ba
\times\,\, S_{||}(p_{||})F^{A}(p_{||};P_{\perp})\gamma^5
\frac{1-i\gamma^1\gamma^2}{2}S_{||}(p_{||}).
\lb{auxfield}
\ea
Then, inserting 
the bilocal field $\tilde{\varphi}(x,y)$ from Eq.(\ref{expansion2}) into 
action (\ref{action11})
and using the BS
equation (\ref{BSequation1}), we obtain the following explicit form 
for the effective action:
$$
S(\tilde{\varphi}) = \sum_{n=2}^{\infty} \frac{i}{n}
\int d^4x_1d^4y_1...d^4x_nd^4y_n
\, \int \frac{d^2P_{\perp}^1\,...\,d^2P_{\perp}^n}{(2\pi)^{2n}}\,
\phi^{A_1}(P^{\perp}_1)\,...\,\phi^{A_n}(P^{\perp}_n)\,
$$
$$
\times\,\, \frac{\mbox{tr}\,
[S_{LLL}^{-1}(x_1,y_1)\chi^{A_1}(y_1,x_2;P^{\perp}_1)\,.\,.\,.\,
S_{LLL}^{-1}(x_{n-1},y_n)\chi^{A_n}(y_n,x_1;P^{\perp}_n)]}{\Pi_{i=1}^n 
\lambda(P_i^{\perp})}\,-\,
\frac{i}{2} \int d^4x_1d^4y_1d^4x_2d^4y_2
\int \frac{d^2P_{\perp}^1d^2P_{\perp}^2}{(2\pi)^4}\, 
$$
\ba
\times\,\, \phi^{A_1}(P^{\perp}_1)\phi^{A_2}(P^{\perp}_2)\,
\frac{\mbox{tr}[
\chi^{A_1}(x_1,y_1;P^{\perp}_1)
S_{LLL}^{-1}(y_1,y_2)\chi^{A_2}(y_2,x_2;P^{\perp}_2)
S_{LLL}^{-1}(x_2,x_1)]}{\lambda(P^{\perp}_1)}.
\lb{action2}
\ea

From the effective action (\ref{action2}), one can obtain the n-point
vertices of NG bosons. In fact, 
using Eq. (\ref{auxfield}), the factorized form of the fermion
propagator (\ref{LLLpropagator}), and the 
fact that
$P(x_{\perp},y_{\perp})$ is a projection operator, we can integrate over
all coordinates in the effective action (\ref{action2}), similarly as it 
was done in the case of the NJL model in Ref. \cite{GM}. 
Then, we get the following expression for the effective action:
$$
S(\phi) =  \sum_{n=2}^{\infty} \Gamma_n,
$$
where the interaction vertices $\Gamma_n$, $n > 2$, are
$$
\Gamma_n = \frac{\pi i|eB|}{2^{n-1} n} \int d^2X_{||} \int 
\frac{d^2k_{||}}{(2\pi )^2}
\int \frac{d^2P_1^{\perp}}{(2\pi )^2}\,...\,\frac{d^2P_n^{\perp}}
{(2\pi )^2}\,
\delta^2(\sum_{i=1}^n \vec{P}_{i}^{\perp})\, 
\phi^{A_1}(P_1^{\perp})\,.\,.\,.\,
\phi^{A_n}(P_n^{\perp})\,
$$
\ba
\times\,\, 
\mbox{tr}\left[S_{||}(k_{||})F^{A_1}(k_{||};P_1^{\perp})\gamma^5
\frac{1-i\gamma^1\gamma^2}{2}\,.\,.\,.\,S_{||}(k_{||})
F^{A_n}(k_{||};P_n^{\perp})
\gamma^5 \frac{1-i\gamma^1\gamma^2}{2}\right] \frac{e^{-\frac{i}{2}\sum_{i<j}
P_i^{\perp} \times P_j^{\perp}}}{\Pi_{i=1}^n \lambda(P_i^{\perp})},
\lb{nvertex}
\ea
and the quadratic part of the action is
$$
\Gamma_2 = -\frac{i|eB|}{16\pi} \int d^2X_{||} \int 
\frac{d^2k_{||}}{(2\pi)^2}
\int \frac{d^2P_{\perp}}{(2\pi )^2}\, 
\frac{\lambda(P_{\perp}) - 1}{\lambda^2(P_{\perp})}\,\phi^{A_1}(P_{\perp})
$$
\ba
\times\,\, \mbox{tr}\left[S_{||}(k_{||})F^{A_1}(k_{||};P_{\perp})\gamma^5
\frac{1-i\gamma^1\gamma^2}{2}S_{||}(k_{||})F^{A_2}(k_{||};-P_{\perp})\gamma^5
\frac{1-i\gamma^1\gamma^2}{2}\right]\,\phi^{A_2}(-P_{\perp}) .
\lb{quadratic}
\ea
For the expressions of the vertices in the case of nonzero longitudinal
momenta, see the Appendix. 

In the next section, we will discuss the connection of the
structure 
of vertices (\ref{nvertex}) with vertices in NCFT.

\section{Type I and Type II nonlocal NCFT}
\label{types}

The derivation of the expressions for vertices (\ref{nvertex}) and 
quadratic part of the action (\ref{quadratic})
is one of the main results of this work (the generalization
of these expressions for the case of nonzero longitudinal momenta
are given in Eqs. (\ref{nvertexl}) and (\ref{quadraticl}) in
the Appendix).
Let us discuss the
connection of the structure of
these vertices with vertices 
in
NCFT. According to \cite{DNS}, an $n$-point vertex in a
noncommutative theory in momentum space has the following
canonical form:
\ba
\int \frac{d^Dk_1}{(2\pi)^D} ... \frac{d^Dk_n}{(2\pi)^D}
\phi(k_1)... \phi(k_n) \delta^D(\sum_i k_i)
e^{-\frac{i}{2} \sum_{i<j} k_i \times k_j},
\label{NCvertex}
\ea
where here $\phi$ denotes a generic field and the exponent
$e^{-\frac{i}{2} \sum_{i<j} k_i \times k_j}\equiv
e^{-\frac{i}{2} \sum_{i<j} k_i \theta k_j}$
is the Moyal exponent
factor. Here the antisymmetric matrix $\theta^{ab}$ determines
the commutator of spatial coordinates:
\ba
[\hat{x}^{a},\hat{x}^{b}] =
i\theta^{ab}\,.
\ea

When can the vertex (\ref{nvertex}) be transformed into the
conventional form (\ref{NCvertex})? In order to answer this question,
it will be convenient to ignore for a moment the fact that 
$F^{A}(p_{||};P_{\perp})$ in (\ref{nvertex})
is a solution of equation (\ref{Aequation}) and consider it as an
arbitrary function of the momenta $p_{||}$ and $P_{\perp}$.
Then,
comparing expressions (\ref{nvertex}) and (\ref{NCvertex}),
it is not difficult to figure out that if the function 
$F(p_{||};P_{\perp})$,
defined as 
\ba
F^{A}(p_{||};P_{\perp}) = (\lambda^A/2)\,F(p_{||};P_{\perp}),
\lb{F1}
\ea
has the factorized form
\ba
F(p_{||};P_{\perp}) = F_{||}(p_{||})F_{\perp}(P_{\perp}),
\lb{factor}
\ea
then there exists a map of the fields $\phi^{A}(P_{\perp})$
into new fields in terms of which
vertices $\Gamma_n$ (\ref{nvertex}) take the conventional
form (\ref{NCvertex}). Indeed, let us introduce new
fields
\ba
\Phi^{A}(P_{\perp}) = \frac{F_{\perp}(P_{\perp})}
{\lambda(P_{\perp})}\phi^{A}(P_{\perp}).
\lb{smear}
\ea
Then, after integrating over $k_{||}$ and taking trace
over Dirac matrices, we get the conventional form for 
$\Gamma_n$:
$$
\Gamma_n = C_{n}|eB|
\int d^2X_{||}
\int \frac{d^2P_1^{\perp}}{(2\pi )^2}\,...\,\frac{d^2P_n^{\perp}}{(2\pi
)^2}\,
\delta^2(\sum_{i=1}^n \vec{P}_{i}^{\perp})
$$
\ba
\times\,\, \mbox{tr}\left[\bar{\Phi}(P_1^{\perp})
\,.\,.\,.\,
\bar{\Phi}(P_n^{\perp})\right]
e^{-\frac{i}{2}\sum_{i<j}
P_i^{\perp} \times P_j^{\perp}}\,\,\,,\,\,\,\,\,\,\,\,\,
\bar{\Phi}\equiv (\lambda^{A}/2)\Phi^{A},
\lb{convertex}
\ea
where $C_{n}$ is some constant.
The propagator of these 
fields is determined from the quadratic part (\ref{quadratic})
of the action:
\ba
\Gamma_2 = C_{2}|eB| \int d^2X_{||} 
\int \frac{d^2P_{\perp}}{(2\pi )^2}\,
(\lambda(P_{\perp}) - 1)\,
\mbox{tr}\left[\bar{\Phi}(P_{\perp})\,\bar{\Phi}(-P_{\perp})\right].
\lb{quadratic1}
\ea
Thus, in terms of
the new fields $\Phi^{A}(P_{\perp})$, vertices (\ref{nvertex}) can
be transformed into the canonical form. As will be shown in the
next section, there exists a special dynamical regime in QED with
a large number of fermion flavors $N_f$ and $M^{2}_{\gamma}
\gtrsim |eB|$ 
in which the constraint (\ref{factor})
can be fulfilled. In fact, this dynamical regime is essentially
the same as that in the NJL model in a strong magnetic field
\cite{GM}. In this case, function $F$ (\ref{F1}) is
$p_{||}$ independent. 
Following Ref. \cite{GM}, the fields $\Phi^{A}(P_{\perp})$ with 
a built-in form factor will be called smeared fields. 

In coordinate space, the interaction 
vertices (\ref{convertex}) of the smeared fields take the form:
\ba
\Gamma_n = \frac{C_{n}}{4\pi^2}|eB|
\int d^2X_{||}\, \int d^2X_{\perp}\, 
\mbox{tr}\left[\bar{\Phi}(X_{\perp})*...
*\bar{\Phi}(X_{\perp})\right],
\lb{coordinate1}
\ea
where the symbol $*$ is the conventional star product
\cite{DNS} relating to the transverse coordinates.
In the space with noncommutative transverse
coordinates $\hat{X}_{\perp}^a$, $a=1,2$, these vertices can be
represented as  
\ba
\Gamma_n = \frac{C_{n}}{4\pi^2}|eB|
\int d^2X_{||} {\bf Tr}\,\,[\hat{\bar{\Phi}}(\hat{X}_{\perp})...
\hat{\bar{\Phi}}(\hat{X}_{\perp})],
\lb{ncspace1}
\ea
where $\hat{\bar{\Phi}}(\hat{X})$ is the Weyl symbol of
the field $\bar{\Phi}(X)$,
the operation ${\bf Tr}$ is defined as in \cite{DNS}, and
\ba
[\hat{X}^{a}_{\perp},\hat{X}^{b}_{\perp}] =
i\frac{1}{eB}\epsilon^{ab} \equiv
i\theta^{ab},\,\,a,b=1,2.
\label{commutator}
\ea
We will refer to theories with a factorized function 
$F(p_{||};P_{\perp})$ as 
type I nonlocal NCFT.

In the case when the function $F(p_{||};P_{\perp})$ in
Eq. (\ref{F1}) is not factorized, one cannot  
represent interaction vertices 
(\ref{nvertex}) in
the form of vertices
of a conventional NCFT. This is the case for $F(p_{||};P_{\perp})$
in the integral equation (\ref{Aequation}) with 
$M^{2}_{\gamma} \ll |eB|$ (see Sec. \ref{QED} below).
However, even in this case,
we still can represent vertex
(\ref{nvertex}) as a nonlocal vertex in the noncommutative
space. Indeed, we can rewrite (\ref{nvertex}) as
\ba
\Gamma_n = \frac{i|eB|}{2^{n+1}\pi n} \int d^4X
\left[\, V_{n}^{A_1...A_n}(-i\nabla^{\perp}_1,...,-i\nabla^{\perp}_n)\, \
\phi^{A_1}(X^{\perp}_1)*...*\phi^{A_n}(X^{\perp}_n)\,\, 
\right]|_{X^{\perp}_1=X^{\perp}_2=...=X^{\perp}},
\lb{coordinate2}
\ea
where the coincidence limit 
$X^{\perp}_1=X^{\perp}_2=...=X^{\perp}$ is taken after 
the action of a
nonlocal operator $V_{n}^{A_1...A_n}$ on the fields $\phi^{A_i}$. 
In momentum space, the
operator $V_{n}^{A_1...A_n}$ is  
$$
V_n^{A_1...A_n}(P^{\perp}_1,...,P^{\perp}_n) = \int 
\frac{d^2k_{||}}{(2\pi)^2}
\mbox{tr}\left[S_{||}(k_{||})F^{A_1}(k_{||};P_1^{\perp})\gamma^5
\frac{1-i\gamma^1\gamma^2}{2}\,.\,.\,.\,S_{||}(k_{||})
F^{A_n}(k_{||};P_n^{\perp})
\gamma^5 \frac{1-i\gamma^1\gamma^2}{2}\right]
$$
\ba
\times \,\,\frac{1}{\Pi_{i=1}^n \lambda(P_i^{\perp})}.
\lb{V}
\ea
By using the fact that the Weyl symbol of the derivative in noncommutative 
space
is given by the operator $\hat{\nabla}_{\perp a}$ 
acting as \cite{DNS}
\ba
\hat{\nabla}_{\perp a} \hat{\phi}(\hat{X}_{\perp}) = -i 
[(\theta^{-1})_{ab}\hat{X}_{\perp}^b,
\hat{\phi}(\hat{X}_{\perp})],
\lb{derivative}
\ea
we obtain the following form for the interaction
vertices in the noncommutative space:
\ba
\Gamma_n = \frac{i|eB|}{2^{n+1}\pi n} \int d^2X_{||} {\bf Tr} \left[\,\{
V_n^{A_1...A_n}(-i\hat{\nabla}^{\perp}_1,...,-i\hat{\nabla}^{\perp}_n)\, \
\hat{\phi}^{A_1}(\hat{X}^{\perp}_1)...\hat{\phi}^{A_n}
(\hat{X}^{\perp}_n)\}_{\hat{X}^{\perp}_1=
\hat{X}^{\perp}_2=...=\hat{X}^{\perp}}\right].
\lb{ncspace2}
\ea

It is clear that NCFT with such vertices are much more 
complicated than 
type I nonlocal NCFT discussed above. We will call them type II nonlocal 
NCFT. As will be shown in Secs. \ref{QED} and \ref{QCD}, QED and
QCD in a strong magnetic field yield examples of such theories. 
For the case of nonzero longitudinal
momentum $P_{||}$, the counterparts of expressions (\ref{coordinate2})
and (\ref{ncspace2}) are derived in the Appendix (see Eqs. 
(\ref{coordinate3}) and (\ref{noncommspace3}) there).

\section{QED with large $N_f$ in a strong magnetic field and NCFT: 
dynamical regime with local interactions}
\label{NJL}

Let us now consider the dynamical regime with such large $N_f$ that 
the $1/N_f$ expansion is reliable, and the coupling 
$\tilde{\alpha_b}$ is so strong that 
$M_{\gamma}^2=2\tilde{\alpha_b}|eB|/\pi$
in (\ref{trunc1}) is of order $|eB|$ or larger. 
In this case,
we have a NJL model with the current-current local interaction,
in which the coupling constant $G =  4\pi \alpha_{b}/M_{\gamma}^2 = 
2\pi^2/{N_f|eB|}$ and  
the ultraviolet cutoff $\Lambda_{||}$ connected with longitudinal momenta 
is $\Lambda_{||}^2 = |eB|$
(see the last paper in Ref. \cite{QED1}).
The dynamical fermion mass is now given by expression (\ref{mass1}) and 
the mass
function $m(k^2)$ is a constant, $m(k^2)=m_{dyn}$.
In this regime,
Eq. (\ref{Aequation}) takes the form
\ba
F(p_{||};P_{\perp}) = \frac{8\pi \alpha_{b} \lambda_{L}(P_{\perp})}
{M_{\gamma}^2} \int 
\frac{d^2u_{\perp}d^2k_{||}}{(2\pi)^4}
\frac{F(k_{||};P_{\perp})}{k_{||}^2+m_{dyn}^2}
e^{iP_{\perp} \times u_{\perp}}\, e^{-\frac{\vec{u}_{\perp}^2}{2|eB|}},
\lb{NJLequation}
\ea
where the function $F(p_{||};P_{\perp})$ is defined in
Eq. (\ref{F1}).
The subscript $L$ in $\lambda_{L}(P_{\perp})$ reflects the 
consideration
of the limit with local interactions here. 
Since the right-hand side of 
equation (\ref{NJLequation}) does not depend on
$p_{||}$, the function $F(p_{||};P_{\perp})$ is  
$p_{||}$-independent, $F(p_{||};P_{\perp}) = F(P_{\perp})$, i.e.,
this dynamics relates to the type I nonlocal NCFT considered in
the previous section. 
Then, 
we immediately find from (\ref{NJLequation}) that 
\ba
\lambda_{L}(P_{\perp}) = \frac{\pi M_{\gamma}^2
e^{\frac{P_{\perp}^2}{2|eB|}}}
{\alpha_{b} |eB|\ln\frac{|eB|}{m_{dyn}^2}} =
\frac{2N_{f}e^{\frac{P_{\perp}^2}{2|eB|}}}{\ln\frac{|eB|}{m_{dyn}^2}},
\lb{NJLlambda}
\ea
where we used $M_{\gamma}^2 = 2\tilde{\alpha}_{b}|eB|/\pi
\gg m_{dyn}^2$.
Using now the on-mass-shell 
condition
$P_{\perp} \to 0$, $\lambda_{L}(P_{\perp}) \to 1$ for the NG bosons 
in equation (\ref{NJLlambda}), 
we arrive at the following gap equation for $m_{dyn}$:
\ba
\frac{2N_f}{\ln\frac{|eB|}{m_{dyn}^2}} = 1.
\lb{gapequation}
\ea
This gap equation yields expression (\ref{mass1}) for the mass. 
It also implies that
$\lambda_{L}$ (\ref{NJLlambda}) can be rewritten in a very simple
form:
\ba
\lambda_{L}(P_{\perp}) = e^{\frac{P_{\perp}^2}{2|eB|}}.
\lb{NJLlambda1}
\ea

The choice of the off-mass-shell
operators $\phi^A$ in expansion
(\ref{expansion2}) is not unique. They are determined by the
choice of their propagator. If one chooses the conventional
composite NG fields $\pi^A = (G/2) \bar{\psi}\gamma_5\lambda^{A} \psi$ 
as $\phi^A$,
their propagator is
\ba
D^{AB}_{\pi} = \frac{8\pi^2\,\,\delta_{AB}}{|eB|\ln\frac{|eB|}
{m_{dyn}^2}
(1 - e^{\frac{-P^2_{\perp}}{2|eB|}})}=
\frac{8\pi^2\,\,\delta_{AB}}{|eB|\ln\frac{|eB|}
{m_{dyn}^2}(1 - \lambda_{L}^{-1}(P_{\perp}))}=
\frac{4\pi^2\,\,\delta_{AB}}
{N_f|eB|(1 - \lambda_{L}^{-1}(P_{\perp}))}.
\lb{NJLpropagator}
\ea
Up to the factor $2$, this propagator coincides with that in the
NJL model in a magnetic field \cite{GM}.
\footnote {In \cite{GM}, the conventional NJL model with the
chiral group $U(1)_{L} \times U(1)_{R}$ and with the number of
fermion colors $N_c$ was considered.
Therefore, comparing these two models, one
should take $N_{c}=1$ and replace the flavor matrices $\lambda^{A}/2$
by $1$. Since $\mbox{tr}(\lambda^{A}/2)^2 = 1/2$, there is an
additional factor $2$ in the propagator of $\pi^A$ in the present model.}

From propagator (\ref{NJLpropagator}) and 
Eq. (\ref{quadratic}) we find
the function $F(P_{\perp})$:
\ba
F(P_{\perp})  = 
2\lambda^{1/2}_{L}(P_{\perp}) = 2
e^{\frac{P_{\perp}^2}{4|eB|}}.
\ea

Interaction vertices (\ref{nvertex}) are nonzero only for even $n$ and
they take now the form
$$
\Gamma_n = -\frac{2|eB|(-1)^{\frac{n}{2}}} {n(n-2)m_{dyn}^{n-2}}
\int d^2X_{||} 
\int \frac{d^2P_1^{\perp}}{(2\pi )^2}\,...\,\frac{d^2P_n^{\perp}}{(2\pi
)^2}\,
\delta^2(\sum_{i=1}^n \vec{P}_{i}^{\perp})
$$
\ba
\times \, \mbox{tr}\left[\bar{\pi}(P_1^{\perp})
\,.\,.\,.\,
\bar{\pi}(P_n^{\perp})\right]
e^{-\frac{\sum_{i}P_{i\perp}^2}{4|eB|}}
e^{-\frac{i}{2}\sum_{i<j}
P_i^{\perp} \times P_j^{\perp}},\,\,\,
\bar{\pi}\equiv (\lambda^{A}/2)\pi^{A}.
\lb{NJLnvertex1}
\ea
These expressions for the vertices agree
with those found in Ref. \cite{GM} (see footnote 6).

Apart from 
the exponentially damping factor
$e^{-\sum_{i}P_{i\perp}^2/4|eB|}$, the form of these
vertices coincide with that in NCFT
with noncommutative space transverse coordinates
$\hat{X}^{a}_{\perp}$ satisfying commutation relation
(\ref{commutator}). The appearance of
the additional Gaussian (form-) factors in vertices 
reflects an inner structure
of composites $\pi^A$ in the LLL
dynamics. These form factors, reflecting the Landau wave functions
on the LLL,
are intimately connected with the holomorphic 
representation in the problem of a free fermion in a strong
magnetic field \cite{FS,GK}. The short-range interactions between 
fermions in this dynamical regime
do not change their Gaussian form.
As was shown
in Ref. \cite{GM}, because of these form factors, the UV/IR mixing is 
absent 
in the model.

As we discussed in the previous section,
in order to take properly into account these
form factors,
it is convenient to introduce new, smeared, fields:
\ba
\bar{\Pi}(X) = e^{\frac{\nabla_{\perp}^2}{4|eB|}}\,\bar{\pi}(X),
\label{smeared}
\ea
where $\nabla_{\perp}^2$ is the transverse Laplacian.
Then, in terms of the smeared fields, the vertices can be rewritten in the
standard form with the Moyal exponent factor only:
$$
\Gamma_n = -\frac{2|eB|(-1)^{\frac{n}{2}}}{n(n-2)m_{dyn}^{n-2}}
\int d^2X_{||}
\int \frac{d^2P_1^{\perp}}{(2\pi )^2}\,...\,\frac{d^2P_n^{\perp}}{(2\pi
)^2}\,
\delta^2(\sum_{i=1}^n \vec{P}_{i}^{\perp})
$$
\ba
\times \, \mbox{tr}\left[\bar{\Pi}(P_1^{\perp})
\,.\,.\,.\,
\bar{\Pi}(P_n^{\perp})\right]
e^{-\frac{i}{2}\sum_{i<j}
P_i^{\perp} \times P_j^{\perp}}.
\lb{NJLnvertex}
\ea
But now the form factor occurs in the
propagator of the smeared fields:
\ba
D^{AB}_{\Pi}(P_{\perp}) =  
e^{\frac{-P^{2}_{\perp}}{2|eB|}}  D^{AB}_{\pi}(P_{\perp}).
\lb{NJLpropagator1}
\ea
In this case, it is again
the form factor $e^{\frac{-P^{2}_{\perp}}{2|eB|}}$, now built in the
propagator $D^{AB}_{\Pi}(P_{\perp})$, that is responsible for the absence
of the UV/IR mixing.

The extension of the present analysis to the case with nonzero
longitudinal momenta $P_{||}$ is straightforward for this
NJL-like dynamical regime (see the Appendix). The results coincide with
those obtained in Ref. \cite{GM}.

Therefore the dynamics in this regime relates to type I
nonlocal NCFT.
As was discussed in the
previous section (see also Ref. \cite{GM}), n-point
vertex (\ref{NJLnvertex}) can be rewritten in the coordinate space
in the standard NCFT form with the star product \cite{DNS}. Moreover,
as was shown in \cite{GM}, in the space with noncommutative 
transverse 
coordinates, one can derive the effective action for the composite fields 
in this model. 
Thus, here we reproduced the results of Ref. \cite{GM} by using the 
method of bilocal
auxiliary
fields.

\section{QED with weak coupling in a strong magnetic field 
and type II nonlocal NCFT}
\label{QED}

In this section, we will consider the dynamics of QED
in a magnetic field in a weak coupling regime, when
the coupling $\tilde{\alpha_b}$ is small (the number of flavors
$N_f$ can now be arbitrary). 
As will become clear in a moment, this dynamics yields an example
of a type II nonlocal NCFT.

According to the analysis in Section II, the integral equation for
$F(p_{||};P_{\perp})$ in this 
regime is
\ba
F(p_{||};P_{\perp}) = 8\pi\alpha_b\lambda_{W}(P_{\perp}) \int 
\frac{d^2u_{\perp}d^2k_{||}}{(2\pi)^4}
\frac{F(k_{||};P_{\perp})}{k_{||}^2+m_{dyn}^2}
\frac{e^{iP_{\perp} \times u_{\perp}} e^{-\frac{\vec{u}_{\perp}^2}{2|eB|}}}
{(k_{||}-p_{||})^2+\vec{u}_{\perp}^2+M_{\gamma}^2},
\lb{QEDequation}
\ea
where $m_{dyn}$ is given in Eq. (\ref{mass}) and
the subscript $W$ in $\lambda_{W}$ reflects the consideration of
the weak coupling regime.
Unlike the integral equation (\ref{NJLequation}), the kernel of this
equation does not have a separable form. Therefore, the function
$F(p_{||};P_{\perp})$ is not factorized in this case and the present
dynamics relates to type II nonlocal NCFT. According to
the analysis in Sec. \ref{types}, its n-point vertices can be 
written 
either through the star product in the form (\ref{coordinate2})
in coordinate space
or in the form (\ref{ncspace2})
in the noncommutative space.

In order to illustrate the difference of this dynamics from that
considered in the previous section, it will be instructive to
analyze it in a special limit with
$P_{\perp}^2 \gg |eB|$. 
\footnote {While in the dynamical regime with the LLL dominance
longitudinal momenta should satisfy the inequality 
$P_{||} \ll \sqrt{|eB|}$, transverse momenta can be large.} 
We will show that in this limit
the approximation with $F(p_{||};P_{\perp})$ being independent
of $p_{||}$ is quite good and, therefore, the dynamics in this limit
can be considered as approximately relating to type I nonlocal NCFT.
However, as will be shown below,
the form factor in this dynamics is very different from
the Gaussian form factor that occurs in the NJL-like dynamics.
This point reflects a long-range character of
the QED interactions.

We start the analysis of integral equation (\ref{QEDequation})
for $P_{\perp}^2 \gg |eB|$
by considering the integral
\ba
I = \int \frac{d^2u_{\perp}}{(2\pi)^2}
\frac{e^{iP_{\perp} \times u_{\perp}} 
e^{-\frac{\vec{u}_{\perp}^2}{2|eB|}}}
{q_{||}^2+\vec{u}_{\perp}^2+M_{\gamma}^2} =
\int \frac{d^2u_{\perp}}{(2\pi)^2}
\frac{e^{i\vec{\Delta}_{\perp}\vec{u}_{\perp}} 
e^{-\frac{\vec{u}_{\perp}^2}{2|eB|}}}
{q_{||}^2+\vec{u}_{\perp}^2+M_{\gamma}^2},
\lb{I1}
\ea
where $q_{||} = k_{||} - p_{||}$, 
$\vec{\Delta}_{\perp}=\frac{\vec{P}_{\perp}}{eB}$
and, for convenience,
we made the change of variable $u_{\perp}^1 \to -u_{\perp}^2$,
$u_{\perp}^2 \to u_{\perp}^1$
in the last
equality.
By representing $e^{-\vec{u}_{\perp}^2/2|eB|}$ and
$(q_{||}^2+\vec{u}_{\perp}^2+M_{\gamma}^2)^{-1}$
through their Fourier transforms, we obtain
\ba
I = \int d^2\Delta_{\perp 1}\, 
\frac{|eB|\,e^{-\frac{|eB|\vec{\Delta}_{\perp 1}^2}{2}}}
{2\pi}\,\,\,
\frac{K_0(|\vec{\Delta}_{\perp} -
\vec{\Delta}_{\perp 1}|\,\sqrt{q_{||}^2+M_{\gamma}^2})}{2\pi},
\lb{involution}
\ea
where $K_0(z)$ is the Bessel function of imaginary argument \cite{GR}.
For $P_{\perp}^2 \gg |eB|$, when $|\vec{\Delta}_{\perp}| \gg 1$, one can 
neglect the
dependence of $K_0$ on $\vec{\Delta}_{\perp 1}$. Then, 
using the asymptotics of $K_0(z)$ at $z \to +\infty$, we find that
\ba
I \approx \frac{1}{2}\left(\frac{|eB|}
{2\pi |P_{\perp}|(q_{||}^2+M_{\gamma}^2)^{1/2}}
\right)^{1/2}
e^{-\frac{|P_{\perp}|(q_{||}^2+M_{\gamma}^2)^{1/2}}{|eB|}},
\lb{I2}
\ea
where $|P_{\perp}| \equiv \sqrt{\vec{P}_{\perp}^2}$.
Since this $I$ as a function of $q_{||}$ exponentially decreases starting 
from
$M_{\gamma}$, it is sufficient 
to take into account only the region with
$q_{||} \alt M_{\gamma}$
and approximate $I$ there by
\ba
I \approx \frac{1}{2}\left(\frac{|eB|}{2\pi 
|P_{\perp}|M_{\gamma}}\right)^{1/2}
e^{-\frac{|P_{\perp}|M_{\gamma}}{|eB|}}.
\lb{I5}
\ea

Thus, we conclude that for $P_{\perp}^2 \gg |eB|$,
a good approximation for equation (\ref{QEDequation}) is
to integrate over $k_{||}^2$ up to $M_{\gamma}^2$ and 
to use 
expression (\ref{I5}) for $I$. This implies that for large $P_{\perp}^2$
the function $F(p_{||};P_{\perp})$ can be taken independent of
$p_{||}$.
Then we find from Eq. (\ref{QEDequation}) 
that
\ba
\lambda_{W}(P_{\perp}) \simeq \frac{1}{\alpha_b\ln\frac{M_{\gamma}^2}
{m_{dyn}^2}} 
\left(\frac{2\pi |P_{\perp}|M_{\gamma}}{|eB|}\right)^{1/2}
e^{\frac{|P_{\perp}|M_{\gamma}}{|eB|}}.
\lb{QEDlambda}
\ea

By choosing $F(P_{\perp}) = 2\lambda_{W}^{1/2}(P_{\perp})$,
we obtain 
the following pion propagator  from Eq.(\ref{quadratic})
(compare with Eq. (\ref{NJLpropagator})):
\ba
D^{AB}(P_{\perp}) = \frac{8\pi^2\delta_{AB}}{|eB|\ln \frac{M^{2}_{\gamma}}
{m_{dyn}^2}
(1-\lambda_{W}^{-1}(P_{\perp}))}.
\label{QEDpropagator}
\ea
And, using Eq. (\ref{nvertex}), we find 
the corresponding vertices:   
$$
\Gamma_n = \frac{2\pi i|eB|}{n}
\int d^2X_{||} \int \frac{d^2k_{||}}{(2\pi )^2}
\int \frac{d^2P_1^{\perp}}{(2\pi )^2}\,...\,\frac{d^2P_n^{\perp}}{(2\pi )^2}\,
\delta^2(\sum_{i=1}^n \vec{P}_{i}^{\perp})
$$
\ba
\times\,\,
\mbox{tr}\left[S_{||}(k_{||})\gamma^5 \bar{\pi}(P_{1}^{\perp})
\frac{1-i\gamma^1\gamma^2}{2}\,.\,.\,.\,S_{||}(k_{||})\bar{\pi}(P_{n}^{\perp})
\gamma^5 \frac{1-i\gamma^1\gamma^2}{2}\right]\, \Pi_{i=1}^{n}\,
\lambda_{W}^{-1/2}(P_i^{\perp})\,
e^{-\frac{i}{2}\sum_{i<j}
P_i^{\perp} \times P_j^{\perp}}.
\lb{QEDnvertex1}
\ea

Now, let us compare expressions (\ref{QEDpropagator}) and (\ref{QEDnvertex1}) 
with their counterparts (\ref{NJLpropagator}) and (\ref{NJLnvertex1})
in the case of local interactions. While the behaviors of 
propagators (\ref{NJLpropagator}) and (\ref{QEDpropagator})
are similar
for large $P^{2}_{\perp} \gg |eB|$ (they both approach a constant as
$P^{2}_{\perp} \to \infty$), the behaviors of vertices 
(\ref{NJLnvertex1}) and (\ref{QEDnvertex1})
are quite different.
While the form factor in vertices (\ref{NJLnvertex1}) has 
the Gaussian form $e^{-\frac{\sum_{i}P_{i\perp}^2}{4|eB|}}$, the form
factor in vertices (\ref{QEDnvertex1}) is proportional to
$(|P_{\perp}|M_{\gamma}/|eB|)^{-1/4}\,\, 
e^{\frac{-|P_{\perp}|M_{\gamma}}{2|eB|}}$
and, therefore, decreases much slower for large $P^{2}_{\perp}$.
The reason of that is a non-local character of the interactions
in QED in a weak coupling limit. To see this more clearly,
let us compare integral equations (\ref{NJLequation}) and
(\ref{QEDequation}). The transition to the local interactions 
corresponds to
the replacement of the propagator 
$[(k_{||}- p_{||})^2+\vec{u}_{\perp}^2+M_{\gamma}^2)]^{-1}$
in Eq. (\ref{QEDequation}) by $M_{\gamma}^{-2}$. This in turn leads
to the replacement of the Bessel 
function
$K_0(|\vec{\Delta}_{\perp}-\vec{\Delta}_{\perp 1}|\,
\sqrt{q_{||}^2+M_{\gamma}^2})$
in Eq. (\ref{involution}) by the delta function $2\pi/M_{\gamma}^2\,\,
\delta^{2}(\vec{\Delta}_{\perp}-\vec{\Delta}_{\perp 1})$. The
substitution of the
delta function in $I$ (\ref{involution}) leads to the Gaussian form 
factor,
which, therefore, is a signature of short-range interactions.
  
\section{Chiral dynamics in QCD in a magnetic field and type II 
nonlocal NCFT}
\label{QCD}

In this section, we will show that the chiral dynamics in QCD in
a strong magnetic field relates to type II nonlocal NCFT. Here under
strong magnetic fields, we understand the fields satisfying 
$|eB| \gg \Lambda^{2}_{QCD}$, where 
$\Lambda_{QCD}$ is the QCD
confinement scale. 

A crucial difference between the dynamics in QED and QCD in
strong magnetic backgrounds is of course the property of asymptotic
freedom and confinement in QCD. The infrared dynamics in quantum
chromodynamics
is much richer and more sophisticated. As was shown in Ref. 
\cite{QCD1},
the confinement scale $\lambda_{QCD}(B)$
in QCD in a strong magnetic field
can be much
less than the confinement scale $\Lambda_{QCD}$ in the vacuum.
As a result,
an anisotropic dynamics of confinement is realized with a rich and
unusual spectrum of very light glueballs.

On the other hand, 
the chiral dynamics in QED and QCD in
strong magnetic backgrounds have a lot in common \cite{QCD1,KLW}.
The point is that the region of momenta relevant for the chiral symmetry 
breaking dynamics is $m_{dyn}^2 \ll
|k^2| \ll |eB|$. If the magnetic field is so strong that the dynamical
fermion mass $m_{dyn}$ is much larger than the confinement scale
$\lambda_{QCD}(B)$, the running coupling $\alpha_{s}$ is small for such
momenta. As a result, the dynamics in that region is essentially 
Abelian. Indeed, while
the contribution of (electrically neutral) gluons
and ghosts in the polarization operator is proportional to
$k^2$, the fermion contribution is proportional to $|eB|$,
similarly to the case of QED in a magnetic field
(see Eq. (\ref{polarization}) and Eq. (\ref{polarization1}))
below). As a result, 
the fermion contribution 
dominates in the relevant region with $|k^2| \ll |eB|$.

Because of the Abelian like structure of the dynamics in this
problem, one can use the results of the analysis in
QED in a magnetic field \cite{QED1}, by introducing appropriate
modifications. One of the modifications is that the chiral symmetry
in QCD in a magnetic field is different from that in QED. Indeed,
since the background magnetic field breaks explicitly the global chiral
symmetry that interchanges the up and down quark flavors (having
different electric charges),
the chiral
symmetry in this problem is $SU(N_{u})_{L}\times SU(N_{u})_{R} \times
SU(N_{d})_{L}\times SU(N_{d})_{R}\times U^{(-)}(1)_{A}$,
where $N_{u}$ and $N_{d}$ are the numbers of up and down quarks,
respectively (the total number of quark flavors is $N_{f}
=N_{u}+N_{d}$). 
The $U^{(-)}(1)_{A}$ is connected with the current which is an
anomaly free linear combination
of the $U^{(d)}(1)_{A}$ and $U^{(u)}(1)_{A}$ currents.
[The $U^{(-)}(1)_{A}$ symmetry is of course absent if either
$N_d$ or $N_u$ is equal to zero].
The generation of quark masses
breaks this symmetry spontaneously down to $SU(N_{u})_{V}\times
SU(N_{d})_{V}$ and, as a result, $N_{u}^{2}+N_{d}^{2}-1$ {\it neutral}
NG bosons occur. 

Another modification is connected with the presence of a new quantum
number, the color. As was shown in Ref. \cite{QCD1}, there exists
a threshold value of the number of colors $N^{thr}_c$
dividing the theories with essentially different dynamics.
For the number of colors $N_c \ll N^{thr}_c$, an anisotropic dynamics of
confinement with the confinement scale $\lambda_{QCD}(B)$
much less than $\Lambda_{QCD}$ and
a rich spectrum of light glueballs is realized. For $N_c$ of order
$N^{thr}_c$ or larger, a conventional confinement dynamics 
with $\lambda_{QCD}(B) \simeq \Lambda_{QCD}$ takes place. 
The threshold value $N^{thr}_c$ grows rapidly with the
magnetic field.
For example, for $\Lambda_{QCD}= 250 \mbox{ MeV}$ and $N_u =1$, $N_d =2$,
the threshold value is
$N^{thr}_c \gtrsim 100$ for $|eB| \gtrsim (1\mbox{
GeV})^2$. We will consider both the case with $N_c \ll N^{thr}_c$
and that with $N_c \gtrsim N^{thr}_c$.

For $N_c \ll N^{thr}_c$,
the dynamical mass $m_{dyn}^{(q)}$
of a $q$-th quark is \cite{QCD1}:
\be
m_{dyn}^{(q)} \simeq C \sqrt{|e_{q}B|}
\left(c_{q}\alpha_{s}\right)^{1/3}
\exp\left[-\frac{2N_{c}\pi}{\alpha_{s} (N_{c}^{2}-1)
\ln(C_{1}/c_{q}\alpha_{s})}\right]
\label{gap}
\ee
(compare with expression (\ref{mass}) for the dynamical mass in QED).
Here $e_{q}$ is the electric charge of the $q$-th quark, 
the numerical factors $C$ and $C_1$ 
are of order one and the constant $c_{q}$ is defined as 
\be
c_{q} = \frac{1}{6\pi}(2N_{u}+N_{d})\left|\frac{e}{e_{q}}\right|.
\ee
The strong coupling $\alpha_{s}$ in Eq. (\ref{gap}) is
related to the scale $\sqrt{|eB|}$, i.e.,
\be
\frac{1}{\alpha_{s}} \simeq b\ln\frac{|eB|}{\Lambda_{QCD}^2},
\quad
b=\frac{11 N_c -2 N_f}{12\pi}.
\label{coupling}
\ee

In QCD, there are two sets of NG bosons related to 
the $SU(N_{u})_{V}$ and $SU(N_{d})_{V}$ symmetries. Their BS
wave functions are defined as in Eq. (\ref{chitilde}) with the
superscript $A$ replaced by $A_u$
($A_d$) for the set connected    
with the $SU(N_{u})_{V}$ ($SU(N_{d})_{V}$). The
corresponding BS equations have the form 
of equation (\ref{f-equation1}) with the coupling $\alpha_b$
replaced by the strong coupling $(N_{c}^2 -1)\alpha_s/2N_c$,
where $(N_{c}^2 -1)/2N_c$ is the quadratic Casimir invariant
in the fundamental representation of $SU(N_c)$. 

Besides these NG bosons, there is one NG boson connected with
the anomaly free $U^{(-)}(1)_A$ discussed above. Its BS wave function
is defined as in
Eqs. (\ref{chitilde}) and (\ref{chitilde1}) but with the
matrix $\lambda^A$ replaced by the traceless matrix
$\tilde{\lambda}^{0}/2 \equiv (\sqrt{N_{d}/N_{f}}\lambda^{0}_{u} -
\sqrt{N_{u}/N_{f}}\lambda^{0}_{d})/2$  \cite{QCD1}. 
Here $\lambda^{0}_{u}$
and $\lambda^{0}_{d}$ are proportional 
to the unit matrices in the up and down flavor sectors, respectively.
They are  normalized as the $\lambda^A$ matrices:
$\mbox{tr}[(\lambda^{0}_u)^2]= \mbox{tr}[(\lambda^{0}_d)^2]= 2$.  

The polarization
operator in the propagator of gluons has the form similar to
that for photons \cite{QCD1}:
\begin{equation}
\Pi \simeq -\frac{\alpha_{s}}{\pi}
\sum_{q=1}^{N_{f}}\frac{|e_{q}B|}{k^{2}_{||}}
\lb{polarization1}
\end{equation}    
(compare with Eq. (\ref{polarization})).
This expression implies that the gluon mass is
\be
M_{g}^2= \sum_{q=1}^{N_{f}}\frac{\alpha_{s}}{\pi}|e_{q}B|=
(2N_{u}+N_{d}) \frac{\alpha_{s}}{3\pi}|eB|.
\label{M_g}
\ee

It is clear that the chiral dynamics
in this case
is similar to that in QED in a weak coupling regime considered
in Sec. \ref{QED} and it relates to the type II nonlocal NCFT.
In the noncommutative space, the expressions
for the vertices of each of the two sets of NG bosons have the form 
(\ref{ncspace2}) ((\ref{noncommspace3}), in the Appendix)
for the case when their fields are independent of   
(depend on) longitudinal coordinates. Notice
that in the leading approximation, the vertices do not mix 
NG fields from the different $SU(N_{u})_{V}$ and $SU(N_{d})_{V}$
sets: the mixing is
suppressed by powers of the small coupling $\alpha_s$
(an analogue of the Zweig-Okubo rule). On the other hand,
the NG boson related to the $U^{(-)}(1)_A$
interacts 
with NG bosons from both these sets without any suppression.

Let us now turn to the case with large $N_c$, in particular, to
the 't Hooft limit $N_c \to \infty$. 
Just a look at expression (\ref{M_g})
for the gluon mass is enough to recognize that the dynamics in this limit
is very different from that considered above. Indeed,
as is well known, the strong coupling constant $\alpha_s$ is proportional
to $1/N_c$ in this limit. More precisely, it rescales as
\be
\alpha_s = \frac{\tilde{\alpha}_s}{N_c},
\label{tilde}
\ee
where the new coupling constant $\tilde{\alpha}_s$ remains finite
as $N_c \to \infty$. Then, expression (\ref{M_g}) implies that
the gluon mass goes to zero in this limit. This in turn implies
that the appropriate approximation is now not
the improved rainbow (ladder) approximation but the 
rainbow (ladder) approximation
itself, when the gluon propagator in the gap (Schwinger-Dyson) equation
and in the BS equation is taken to be bare with $\Pi = 0$
\cite{QCD1}. In other words, gluons are massless and
genuine long range interactions
take place in this regime. 
The dynamical mass of quarks now is
\be
m^{(q)}_{dyn} = C \sqrt{|e_{q}B|}
\exp\left[-{\pi}
\left(\frac{\pi}{\tilde{4\alpha}_s}\right)^{1/2}
\right],
\label{m{infty}}
\ee
where the constant $C$ is of order one. 
As was shown in Ref. \cite{QCD1}, this expression is
a good approximation for the quark mass when
$N_c$ is of order $N^{thr}_c$ or larger.
As to the BS equation, repeating the analysis of Sec. 
\ref{QED},
it is easy to show that, unlike the QED case, 
the amplitude $F(p_{||},P_{\perp})$, defined in Sec. \ref{types},
is not factorized in this dynamical regime
even for large transverse momenta $P_{\perp}^2 \gg |eB|$. Therefore,
the dynamics with large $N_c$ yields even a more striking example of 
the type 
II
nonlocal NCFT than the previous dynamical regime with 
$N_c \ll N^{thr}_c$.  

\section{More about the connection between field theories in 
a magnetic field and NCFT}
\label{general}

What are the origins 
of the connection between field theories in a magnetic field and NCFT?
As was emphasized in Ref. \cite{GM}
and in Sec. \ref{2} of this paper, it is the Schwinger phase
in the LLL fermion propagator and BS wave functions of
neutral composites
that leads to the Moyal factor (a
signature of NCFT) in interaction vertices of neutral composites.
But what is the origin of the Schwinger phase itself? Let us
show that it reflects the existence of the group of
magnetic translations in an external magnetic field 
\cite{GMS2,translations}. The generators of this group in
the symmetric gauge, used in this paper, are:
\begin{equation}
\hat P_{1}=\frac{1}{i}\frac{\partial}{\partial x_1}-
\frac{\hat{Q}}{2} Bx_2~, \quad
\hat{P}_{2} = \frac{1}{i}\frac{\partial}{\partial x_2} +
\frac{\hat{Q}}{2} Bx_1~, \quad
\hat P_{3}=\frac{1}{i}\frac{\partial}{\partial x_3} ,  
\label{generators}
\end{equation}
where $\hat{Q}$ is the charge operator. The commutators of these
generators are:
\begin{equation}
\bigl[\hat P_{1},\hat P_{2}\bigr] = \frac{1}{i}
\hat{Q}B~, \quad
\bigl[\hat P_{1},\hat P_{3}\bigr] =
\bigl[\hat P_{2},\hat P_{3}\bigr] = 0 . 
\label{commutatorM}
\end{equation}
Therefore all the commutators equal zero for neutral states, and
the momentum $\vec{P}=(P_1,P_2,P_3)$ 
of their center of mass is a good quantum number. 

It is easy to check that
the structure of the generators (\ref{generators}) implies the presence
of the Schwinger phase in the matrix elements of the time ordered
bilocal operator $T(\psi(x) \bar{\psi}(y))$ taken between two 
arbitrary neutral states, $|a;P_a>$ and $|b;P_b>$. More precisely,
\ba
M_{ba}(x,y;P_b,P_a) \equiv 
<P_b;b|T\psi(x)\bar{\psi}(y)|a;P_a> = e^{-i(P_a - P_b)X}
e^{ier^{\mu}A_{\mu}^{ext}(X)} \tilde{M}_{ba}(r;P_b,P_a),
\lb{M}
\ea
where $r=x-y$, $X=\frac{x+y}{2}$ (compare with Eq. (\ref{chitilde})).
The second exponent factor on the right hand side of this equation
is the Schwinger phase. Taking $|a;P_a>$ and $|b;P_b>$ to be
the vacuum state $|0>$, we get the fermion propagator. And taking
$|b;P_b> = |0>$ and $|a;P_a>$ to be a state of some 
neutral composite, we get the BS wave function of this
composite. 
Thus, the group of magnetic translations is
in the heart of the connection between the field dynamics in a
magnetic field and NCFT. 
In particular, one of the consequences of this
consideration is that although the treatment of neutral composites other 
than NG bosons is much more involved, one can be sure that, in a
strong magnetic field, the 
noncommutative structure of interaction vertices 
which include those composites is similar to that 
of the vertices for NG bosons that was
derived in Secs. \ref{2}, \ref{types}, and in the Appendix.   

Now, what is the form of interaction vertices which include such
{\it elementary} electrically neutral fields as photon and gluon
fields in QED and QCD, respectively. This problem in QED has been
recently considered in Ref. \cite{GHM}. As was shown there, in the
LLL approximation, fermion loops infect 
a noncommutative structure for $n$-point photon vertices
and, as a result,
the Moyal factor occurs there. The situation in QCD is more subtle.
In that case, besides induced gluon vertices generated by quark
loops, there are also triple and quartic gluon vertices in the initial
QCD action. There is of course no Moyal factor in those vertices.
Still, since the Abelian approximation is reliable in the description
of the chiral dynamics in a strong magnetic field in QCD (see the 
discussion in the previous section), the situation 
for this dynamics in QCD is similar
to that in QED. However, the description of
the confinement dynamics in QCD, relating
to the deep infrared region, is much more involved, although the influence
of the magnetic field on that dynamics 
is quite essential \cite{QCD1}.   

The last point we want to address in this section
is the reliability of the LLL
approximation in a strong magnetic field. As to the chiral dynamics,
there are solid arguments that it is a reliable approximation for 
it
\cite{GMS1,GMS2,QED1}. In particular, its reliability
was shown explicitly in
the NJL model in a magnetic field in the leading order in $1/N_{c}$
expansion \cite{GMS2}. On the other hand, as has been recently shown 
in Ref. \cite{GHM}, the cumulative effect of higher Landau levels
can be important for n-point photon vertices, at least in some
kinematic regions (a nondecoupling phenomenon). It is however
noticeable that in the kinematic region with momenta $k_{i\perp}^2 
\gg |k_{i\parallel}^2|$, which provides the dominant contribution
in the chiral dynamics \cite{QED1}, the LLL contribution is
dominant \cite{GHM}. Thus, although this question deserves
further study, the assumption about the LLL dominance in the chiral 
dynamics in relativistic field theories 
seems to be well justified. 
  
\section{Conclusion}
\label{conclusion}

The main result of this paper is that
the chiral dynamics in 
QED and QCD in a 
strong magnetic field 
determine complicated nonlocal NCFT (type II nonlocal NCFT).
These NCFT are quite different both from the NCFT considered
in the literature and  
the NCFT corresponding to the NJL model in a magnetic field
(type I nonlocal NCFT)
\cite{GM}. While in type I NCFT there exists a field transformation
that puts interaction vertices in the conventional form 
(with a cost of introducing an exponentially damping form factor
in field propagators),
no such a transformation exists for type II nonlocal NCFT. 

The reason 
of this distinction between the two types of models is
in the characters of their interactions, being short-range in
the NJL-like models and long-range in gauge theories. While
the influence of the short-range interactions
on the LLL dynamics is quite minor,
the long-range interactions change essentially
that dynamics. As a result, the structure of neutral composites and
manifestations of 
nonlocality 
in gauge theories in strong magnetic backgrounds are much richer. 

We believe that both these types of 
nonlocal NCFT can be relevant not only for relativistic field theories
but also for nonrelativistic systems in a magnetic field.
In particular, while type I NCFT can be relevant for the
description of the quantum Hall effect in condensed matter systems
with short-range interactions \cite{IKS,CTZ,Hall},
type II NCFT can be relevant in studies of this effect in
condensed
matter systems with long-range interactions. Concrete examples of
such systems are provided by
carbon materials, in particular, by highly oriented pyrolytic
graphite (HOPG), 
where Coulomb-like interactions take place \cite{GGV,Khv,GGMS}.
The effective theory of this system is QED$_{2+1}$. Recent
experiments in HOPG in strong magnetic fields \cite{Kop}
suggest the
existence of the quantum Hall effect in this system. It would be
interesting to clarify the dynamics of this effect in the
framework of the QED$_{2+1}$ effective theory.

As is well known,
dynamics and symmetry structures  
of field theories in two spatial dimensions 
can significantly differ from those in three spatial 
dimensions.
These differences can be especially important in gauge 
theories.
The reason of that is 
a possibility of generating some unique 2+1-dimensional
terms, like the Chern-Simons term, in effective actions of
gauge theories. This makes the studies of gauge
theories in a magnetic field in 2+1 dimensions to be interesting not only 
for applications in condensed matter physics but also from the
theoretical point of view.
We are planning to study $2+1$
dimensional gauge theories in a magnetic field
by using the NCFT approach elsewhere.
[In the NCFT approach, some aspects
of dynamics in the $2+1$ dimensional NJL model in a magnetic field were
considered in Ref. \cite{GM}].

Another potentially interesting problem would be 
the examination of the existence 
of type II nonlocal
NCFT in string
theories in magnetic backgrounds with broken supersymmetries.   

\begin{acknowledgments}
We thank Michio Hashimoto and Igor Shovkovy for useful remarks.
The work was supported by the Natural 
Sciences and Engineering Research Council of Canada.

\end{acknowledgments}

\vspace{8mm}

\centerline{\bf Appendix}
\vspace{5mm}

In the main body of the paper we restricted our analysis to the case when
fields $\phi^A(X)$ depend only on the transverse coordinates $X_{\perp}$.
In this Appendix,
we consider the general case of fields $\phi^A(X)$ depending on both
transverse and longitudinal coordinates.

Instead expression (\ref{expansion2}), now we have
the following representation for the bilocal field $\tilde{\varphi}(x,y)$: 
\ba
\tilde{\varphi}(x,y)= \int \frac{d^4P}{(2\pi)^4} \phi^{A}(P)
\chi^{A}(x,y;P),
\lb{expansion3}
\ea
where the structure of the BS wave function
$\chi^A(x,y;P)$ is described in Eqs. (\ref{chitilde}), (\ref{tildechi}) 
and
(\ref{f-equation2}). While for the case $P_{||}=0$ the effective action 
was given in expression (\ref{action2}), it now takes 
the form
$$
S(\tilde{\varphi}) = \sum_{n=2}^{\infty} \frac{i}{n} \int 
d^4x_1d^4y_1...d^4x_nd^4y_n
\, \int \frac{d^4P_1\,...\,d^4P_n}{(2\pi)^{4n}}\,
\phi^{A_1}(P_1)\,...\,\phi^{A_n}(P_n)\,
$$
$$
\times\,\, \frac{\mbox{tr}\,
[S_{LLL}^{-1}(x_1,y_1)\chi^{A_1}(y_1,x_2;P_1)\,.\,.\,.\,
S_{LLL}^{-1}(x_{n-1},y_n)\chi^{A_n}(y_n,x_1;P_n)]}{\Pi_{i=1}^n
\lambda(P_i)}\,-\,
\frac{i}{2} \int d^4x_1d^4y_1d^4x_2d^4y_2
\int \frac{d^4P_1d^4P_2}{(2\pi)^8}\,
$$
\ba
\times\,\, \phi^{A_1}(P_1)\phi^{A_2}(P_2)\,
\frac{\mbox{tr}[
\chi^{A_1}(x_1,y_1;P_1)S_{LLL}^{-1}(y_1,y_2)\chi^{A_2}(y_2,x_2;P_2)
S_{LLL}^{-1}(x_2,x_1)]}{\lambda(P_1)}.
\lb{action3}
\ea

It is convenient to represent $f^A(p_{||};P)$, defined in
Eq. (\ref{f-equation2}), as
\ba
f^A(p_{||};P) = S_{||}(p_{||}+\frac{P_{||}}{2}) G^A(p_{||};P)
S_{||}(p_{||}-\frac{P_{||}}{2}).
\lb{G}
\ea
[Note that a $\gamma$-matrix structure of
$G^A(p_{||};P)$ is determined from the corresponding BS
equation and it can be
different from that in
Eq. (\ref{pion}).] 
Then using (\ref{chitilde}) and (\ref{tildechi}) and integrating over
$p_{\perp}$, we get 
$$
\chi^A(x,y;P)= P(x_{\perp},y_{\perp}) \int \frac{d^2p_{||}}
{2(2\pi)^2} e^{-iP\frac{x+y}{2}}
e^{-ip_{||}(x^{||}-y^{||})}
e^{-\frac{\vec{P}_{\perp}^2}{4|eB|}}
e^{\frac{\epsilon^{ab}P_{\perp}^a(x_{\perp}^b-y_{\perp}^b) 
\mbox{sign}(eB)}{2}}
\,
$$
\ba
\times\,\, S_{||}(p_{||}+\frac{P_{||}}{2})G^{A}(p_{||};P)
S_{||}(p_{||}-\frac{P_{||}}{2})
\lb{auxfield1}
\ea
(compare with Eq.(\ref{auxfield})). Substituting $\chi^A(x,y;P)$ in
Eq. (\ref{action3}), we obtain the effective action in momentum space:
$$
S(\phi)= \sum_{n=2}^{\infty} \Gamma_n,
$$
where the interaction vertices $\Gamma_n$, $n > 2$, are
$$
\Gamma_n = \frac{2^{3-n}\pi^3 i|eB|}{n} \int
\frac{d^2k_{||}}{(2\pi )^2}
\int \frac{d^4P_1}{(2\pi )^4}\,...\,\frac{d^4P_n}
{(2\pi )^4}\,
\delta^4(\sum_{i=1}^n P_{i})\,
\phi^{A_1}(P_1)\,.\,.\,.\,
\phi^{A_n}(P_n)\,
$$
$$
\times\,\,
\mbox{tr} 
[G^{A_1}(k_{||} - \frac{P_1^{||}}{2}; P_1) 
S_{||}(k_{||} - P_1^{||})
 G^{A_2}(k_{||} - P_1^{||} - \frac{P_2^{||}}{2}; P_2)
 S_{||}(k_{||} - P_1^{||} - P_2^{||}) 
$$
\ba
...G^{A_n}(k_{||} - \sum_{i=1}^{n-1} P_i^{||} - \frac{P_n^{||}}{2}; P_n)
S_{||}(k_{||} - \sum_{i=1}^{n} P_i^{||})]\,\,
\frac{e^{-\frac{i}{2}\sum_{i<j}
P_i^{\perp} \times P_j^{\perp}}}{\Pi_{i=1}^n \lambda(P_i)}
\lb{nvertexl}
\ea
(compare with expression (\ref{nvertex})),
and the quadratic part of the action is
$$
\Gamma_2 =
-\frac{\pi i|eB|}{4} \int \frac{d^2k_{||}}{(2\pi )^2}
\int \frac{d^4P}{(2\pi )^4}\,
\frac{\lambda(P) - 1}{\lambda^2(P)}\,\phi^{A_1}(P)
$$
\ba
\times\,
\mbox{tr} 
\left[G^{A_1}(k_{||} - \frac{P_{||}}{2}; P) S_{||}(k_{||} - P_{||})
 G^{A_2}(k_{||} - \frac{P_{||}}{2}; - P) S_{||}(k_{||})\right]
\,\phi^{A_2}(-P)
\lb{quadraticl}
\ea
(compare with expression (\ref{quadratic})).

Further, in coordinate space, the
interaction vertices take the form similar to that in
Eq. (\ref{coordinate2}):
\ba
\Gamma_n = \frac{i|eB|}{2^{n+1}\pi n} \int d^4X
\left[\, V_{n}^{A_1...A_n}(-i\nabla_1,...,-i\nabla_n)\, \
\phi^{A_1}(X_1)*...*\phi^{A_n}(X_n)\,\, \right]|_{X_1=X_2=...=X}\,\,,
\lb{coordinate3}
\ea
where the nonlocal operator $V_{n}^{A_1...A_n}$ now depends both on 
transverse $\nabla_{\perp}$ and longitudinal $\nabla_{||}$.
In momentum space, this operator is
$$
V_n^{A_1...A_n}(P_1,...,P_n) = \int
\frac{d^2k_{||}}{(2\pi)^2}
\mbox{tr}
[G^{A_1}(k_{||} - \frac{P_1^{||}}{2}; P_1)
S_{||}(k_{||} - P_1^{||})
 G^{A_2}(k_{||} - P_1^{||} - \frac{P_2^{||}}{2}; P_2)
 S_{||}(k_{||} - P_1^{||} - P_2^{||}) 
$$
\ba
...G^{A_n}(k_{||} - \sum_{i=1}^{n-1} P_i^{||} - \frac{P_n^{||}}{2}; P_n)
S_{||}(k_{||} - \sum_{i=1}^{n} P_i^{||})]
\,\,\frac{1}{\Pi_{i=1}^n \lambda(P_i)}
\ea
(compare with Eq. (\ref{V})).
Finally, the interaction vertices are given by the following expression in 
noncommutative space:
$$
\Gamma_n = \frac{i|eB|}{2^{n+1}\pi n} \int d^2X_{||}\,\, \mbox{\bf Tr}
\,\,[\,\{V_{n}^{A_1...A_n}(-i\nabla_1^{||}, -i\hat{\nabla}_1^{\perp},...,
-i\nabla_n, -i\hat{\nabla}_n^{\perp})\, \
\phi^{A_1}(X_1^{||},\hat{X}_1^{\perp})
$$
\ba
...\phi^{A_n}(X_n^{||},\hat{X}^{\perp})
\,\,\}_{X_i^{||}=X^{||},\,\,\hat{X}_i^{\perp}=\hat{X}^{\perp}}]
\lb{noncommspace3}
\ea
where here the subscript
$i$ runs from 1 to $n$ (compare with expression (\ref{ncspace2})).

What is the connection between the forms of the function
$f^{A}(p_{||};P)$ in the cases with zero
and nonzero longitudinal momentum $P_{||}$ ? In regard to this question, 
it is appropriate to recall
the Pagels-Stokar (PS) approximation \cite{PS}
for a BS wave function $\chi^{A}(p;P)$, which is often used
in Lorentz invariant
field theories. It is assumed in this approximation
that the amputated BS wave function,
defined as
\ba
\hat{\chi}^{A}(p;P) \equiv S^{-1}(p +\frac{P}{2})\chi^{A}(p;P)
S^{-1}(p -\frac{P}{2}),
\ea
is approximately the same for the cases with zero and nonzero
$P$, i.e., 
$\hat{\chi}^{A}(p;P) \simeq \hat{\chi}^{A}(p)$ where
$\hat{\chi}^{A}(p) \equiv \hat{\chi}^{A}(p;P)|_{P=0}$. 
Then, in this approximation,
the BS wave function $\chi^{A}(p;P)$ is 
\ba
\chi^{A}(p;P) = S(p +\frac{P}{2})\hat{\chi}^{A}(p)
S(p -\frac{P}{2}),
\ea
i.e., the whole dependence of the BS wave function on
the momentum $P$ comes from the fermion propagator. 
It is known (for a review, see 
Ref. \cite{Mir}) that the PS approximation can be justified
both for weak coupling dynamics and, in the case of the NJL model,
in the regime with large $N_c$. 

Here we would like to suggest an
anisotropic version of the PS approximation for dynamics
in a magnetic field. The main assumption we make is that the function
$G^A(p_{||};P)$, defined in Eq. (\ref{G}), is approximately
$P_{||}$ independent, i.e.,
\ba
G^A(p_{||};P) \simeq F^{A}(p_{||};P_{\perp})\gamma^5 
\frac{1-i\gamma^1\gamma^2}{2},   
\lb{G1}
\ea
where the function $F^{A}(p_{||};P_{\perp})$ is defined in Eq. 
(\ref{pion}).
Then, the whole dependence of the function 
$f^{A}(p_{||},P)$ (\ref{G})
on
the longitudinal momentum $P_{||}$ comes from the fermion propagator:
\ba
f^A(p_{||};P) = S_{||}(p_{||}+\frac{P_{||}}{2}) 
F^{A}(p_{||};P_{\perp})\gamma^5 \frac{1-i\gamma^1\gamma^2}{2}
S_{||}(p_{||}-\frac{P_{||}}{2})
\lb{G2}
\ea
(compare with Eq. (\ref{pion})).

We utilized expression (\ref{G1}) in QED in a magnetic field in the
dynamical regime with the local interaction considered in Sec. \ref{NJL}. 
In that case, the function $F^{A}(p_{||};P_{\perp})$ is a constant. Then,
using expression (\ref{nvertexl}) for $\Gamma_n$ with
$G^A$ in Eq. (\ref{G1}), 
it is not difficult
to derive the n-point vertices for fields $\phi^{A}(P)$ for
a general momentum $P$. The result coincides with that obtained in
Ref. \cite{GM} in the NJL model.

Expressions (\ref{G1}) and (\ref{G2}) can also be useful for the 
analysis 
of the dynamics in
QED and QCD in a magnetic field in the weak coupling regime. 
As was shown in Secs. \ref{QED} and \ref{QCD}, in those
cases the function $F^{A}(p_{||};P_{\perp})$ depends 
on both momenta $p_{||}$ and $P_{\perp}$ (the dynamics in this regime 
relate to 
type
II NCFT). The determination of this dependence is a nontrivial problem and 
is outside the scope of this paper.


\vspace{8mm}


\begin{thebibliography}{}

\bibitem{DNS} M. R. Douglas and N. A. Nekrasov,
Rev.\ Mod.\ Phys.\ {\bf 73}, 977 (2001);
R. J. Szabo,
Phys.\ Rep.\ {\bf 378}, 207 (2003).

\bibitem{DJT} G.V. Dunne, R. Jackiw and C. Trugenberger,
 Phys.\ Rev.\ D {\bf 41}, 661 (1990).

\bibitem{BS} D. Bigatti and L. Susskind, Phys.\ Rev.\ D {\bf 62}, 066004
(2000).

\bibitem{IKS} S. Iso, D. Karabali and B. Sakita,
Nucl.\ Phys.\  B {\bf 388}, 700 (1992);
Phys.\ Lett.\ B  {\bf 296}, 143 (1992).

\bibitem{CTZ}
A. Cappelli, C. A. Trugenberger and G. R. Zemba,
Nucl.\ Phys.\  B {\bf 396}, 465 (1993).

\bibitem{GJPP} Z. Guralnik, R. Jackiw, S. Y. Pi and
A. P. Polychronakos,
Phys.\ Lett.\ B  {\bf 517}, 450 (2001);

\bibitem{strings} A. Connes, M.R. Douglas, and A. Schwarz,
JHEP {\bf 9802}, 003 (1998);
M.R. Douglas and C. Hull, JHEP {\bf 9802}, 008 (1998).

\bibitem{SW}
N. Seiberg and E. Witten, JHEP {\bf 9909}, 032 (1999).

\bibitem{GM} E.~V.~Gorbar and V.~A.~Miransky, Phys.\ Rev.\ D {\bf 70},
105007 (2004).

\bibitem{MRS} S. Minwalla, M. Van Raamsdonk and N. Seiberg,
JHEP {\bf 0002}, 020 (2000).

\bibitem{FS} L. D. Faddeev and A. A. Slavnov, {\it Gauge Fields:
Introduction to Quantum Theory} (Addison-
Wisley, Redwood City, Calif., 1991)

\bibitem{GK} S. Girvin and T. Jach, Phys. Rev. {\bf B29}, 5617 (1984);
S. Kivelson, C. Kallin, D. Arovas, and J. Schrieffer, Phys.Rev.
{\bf B36}, 1620 (1987);
G.V. Dunne and R. Jackiw, Nucl.\ Phys.\ Proc.\ Suppl.\ C {\bf 33}, 114
(1993).

\bibitem{Hall}
L. Susskind, hep-th/0101029;
A. P. Polychronakos, JHEP 0104 (2001) 011;
A. Cappelli and M. Riccardi, hep-th/0410151.

\bibitem{GMS1}
V.~P.~Gusynin, V.~A.~Miransky, and I.~A.~Shovkovy,
Phys.\ Rev.\ Lett.\  {\bf 73}, 3499 (1994);
Phys.\ Rev.\  {\bf D52}, 4718 (1995);
Phys.\ Lett.\  B {\bf 349}, 477 (1995).

\bibitem{GMS2} V.~P.~Gusynin, V.~A.~Miransky and I.~A.~Shovkovy,
Nucl.\ Phys.\  B {\bf 462}, 249 (1996).

\bibitem{QED1}
V.~P.~Gusynin, V.~A.~Miransky, and I.~A.~Shovkovy,
Phys.\ Rev.\ Lett.\  {\bf 83}, 1291 (1999);
Nucl.\ Phys.\ B {\bf 563}, 361 (1999);
V.~P.~Gusynin, V.~A.~Miransky, and I.~A.~Shovkovy,
Phys.\ Rev.\ D {\bf 67}, 107703 (2003).

\bibitem{QED2}
V.~P.~Gusynin, V.~A.~Miransky, and I.~A.~Shovkovy,
Phys.\ Rev.\ D {\bf 52}, 4747 (1995);
C.~N.~Leung, Y.~J.~Ng, and A.~W.~Ackley,
Phys.\ Rev.\ D {\bf 54}, 4181 (1996);
D.~S.~Lee, C.~N.~Leung, and Y.~J.~Ng,
Phys.\ Rev.\ D {\bf 55}, 6504 (1997);
{\em ibid.} D {\bf 57}, 5224 (1998);
D.~K.~Hong, Y.~Kim, and S.~J.~Sin,
Phys.\ Rev.\ D {\bf 54}, 7879 (1996);
E.~J.~Ferrer and V.~de la Incera,
Phys.\ Rev.\ D {\bf 58}, 065008 (1998);
V.~P.~Gusynin and A.~V.~Smilga,
Phys.\ Lett.\ B {\bf 450}, 267 (1999).

\bibitem{QCD1}  V.~A.~Miransky and I.~A.~Shovkovy,
Phys.\ Rev.\ D {\bf 66}, 045006 (2002).

\bibitem{QCD2} S.~Schramm, B.~Muller, and A.~J.~Schramm,
Mod.\ Phys.\ Lett.\ A {\bf 7}, 973 (1992);
I.~A.~Shushpanov and A.~V.~Smilga,
Phys.\ Lett.\ B {\bf 402}, 351 (1997);
N.~O.~Agasian and I.~A.~Shushpanov,
Phys.\ Lett.\ B {\bf 472}, 143 (2000);
V.~C.~Zhukovsky, V.~V.~Khudyakov, K.~G.~Klimenko, and D.~Ebert,
JETP Lett.\  {\bf 74}, 523 (2001)
[Pisma Zh.\ Eksp.\ Teor.\ Fiz.\  {\bf 74}, 595 (2001)];
D.~Ebert, V.~V.~Khudyakov, V.~C.~Zhukovsky, and K.~G.~Klimenko,
Phys.\ Rev.\ D {\bf 65}, 054024 (2002).

\bibitem{KLW}
D.~Kabat, K.~Lee, and E.~Weinberg,
Phys.\ Rev.\ D {\bf 66}, 014004 (2002).

\bibitem{Kl} H.~Kleinert,
Phys.\ Lett.\ B {\bf 62}, 429 (1976);
E.~Shrauner,
Phys.\ Rev.\ D {\bf 16}, 1887 (1977).

\bibitem{K} T.~Kugo,
Phys.\ Lett.\ B {\bf 76}, 625 (1978).

\bibitem{Schwinger} J. Schwinger, Phys.\ Rev.\ {\bf 82}, 664 (1951).

\bibitem{QED3}
J.~Alexandre, K.~Farakos and G.~Koutsoumbas,
Phys.\ Rev.\ D {\bf 64}, 067702 (2001);
E. Elizalde, E.~J.~Ferrer and V.~de la Incera,
Phys.\ Rev.\ D {\bf 68}, 096004 (2003).

\bibitem{GR} I. S. Gradshteyn and I. M. Ryzhik, {\sl Table of Integrals, 
Series
and Products} (Academic Press, Orlando, 1980).

\bibitem{translations} J. Zak, Phys. Rev. {\bf 134}, A1602 (1964);
J. E. Avron, I. W. Herbst, and B. Simon, Ann. Phys. {\bf 114},
431 (1978).

\bibitem{GHM} E.V. Gorbar, M. Hashimoto, and V. A. Miransky,
Phys.\ Lett.\ B {\bf 611}, 207 (2005).

\bibitem{GGV}  J. Gonz\'{a}lez, F. Guinea, and
M. A. H. Vozmediano, Phys. Rev. B {\bf 59}, R2474 (1999);
Phys. Rev. B {\bf 63}, 134421 (2001).

\bibitem{Khv} D. V. Khveshchenko,
Phys. Rev. Lett. {\bf 87}, 206401 (2001);
{\em ibid.} {\bf 87}, 246802 (2001).

\bibitem{GGMS}
E.~V.~Gorbar, V.~P.~Gusynin, V.~A.~Miransky, and I.~A.~Shovkovy,
Phys. Rev. B {\bf 66}, 045108 (2002); 
Phys. Lett. A {\bf 313}, 472 (2003).

\bibitem{Kop} Y. Kopelevich et al., Phys. Rev. Lett. {\bf 90}, 156402 
(2003).

\bibitem{PS} H.~Pagels and S.~Stokar,
Phys.\ Rev.\ D {\bf 20}, 2947 (1979).

\bibitem{Mir} V. A. Miransky, {\it Dynamical Symmetry Breaking in 
Quantum Field Theories} (World Scientific Co., Singapore, 1993).


\end{thebibliography}
\end{document}